\documentclass[a4paper,twocolumn,11pt,accepted=2021-05-10]{quantumarticle}
\pdfoutput=1
\usepackage[utf8]{inputenc}
\usepackage[english]{babel}
\usepackage[T1]{fontenc}
\usepackage{amsmath}
\usepackage{hyperref}
\usepackage{amssymb}
\usepackage{tikz}
\usepackage{lipsum}
\usepackage{lipsum}
\usepackage{xcolor}
\usepackage{graphicx}% Include figure files
\usepackage{bm}% bold math
\usepackage{siunitx}
\usepackage{xcolor}
\usepackage{dsfont}
\usepackage{tikz}
\usepackage{tikz-3dplot}
\usepackage{qcircuit}
\usepackage{hyperref}
\usepackage[numbers,sort&compress]{natbib}

\begin{document}

\title{Algorithmic Error Mitigation Scheme for Current Quantum Processors}

\author{Philippe Suchsland}
 \affiliation{%
Institute for Theoretical Physics, ETH Zurich, 8093 Zurich, Switzerland
 }%
  \affiliation{%
 IBM Quantum, IBM Research -- Zurich, 8803 Rueschlikon, Switzerland
}%
 \affiliation{%
 Department of Physics, University of Zurich, Winterthurerstrasse 190, 8057 Zurich, Switzerland
 }%
 \author{Francesco Tacchino}
\affiliation{%
 IBM Quantum, IBM Research -- Zurich, 8803 Rueschlikon, Switzerland
}%
\author{Mark H.\ Fischer}%
\affiliation{%
 Department of Physics, University of Zurich, Winterthurerstrasse 190, 8057 Zurich, Switzerland
}%
\author{Titus Neupert}
\affiliation{%
 Department of Physics, University of Zurich, Winterthurerstrasse 190, 8057 Zurich, Switzerland
}%
\author{Panagiotis Kl.\ Barkoutsos}
\affiliation{%
 IBM Quantum, IBM Research -- Zurich, 8803 Rueschlikon, Switzerland
}%
\author{Ivano Tavernelli}
\email{ita@zurich.ibm.com}%
\affiliation{%
 IBM Quantum, IBM Research -- Zurich, 8803 Rueschlikon, Switzerland
}

\maketitle

\begin{abstract}
We present a hardware agnostic error mitigation algorithm for near term quantum processors inspired by the classical Lanczos method. This technique can reduce the impact of different sources of noise at the sole cost of an increase in the number of measurements to be performed on the target quantum circuit, without additional experimental overhead. We demonstrate through numerical simulations and experiments on IBM Quantum hardware that the proposed scheme significantly increases the accuracy of cost functions evaluations within the framework of variational quantum algorithms, thus leading to improved ground state calculations for quantum chemistry and physics problems beyond state-of-the-art results.
\end{abstract}

\textit{Introduction.}
%%%%%%%%%%%%%%%%%%%%%%%%%%%%%%%%%%%%%%%%%%%%%%%%%%%%%%%%%%%%%%%%
One of the main limitations of available quantum computers is the sensitivity to noise. Applications like quantum chemistry, machine learning and finance require significant precision in the calculation of observables or classical cost functions. However, current hardware noise levels only allow for limited accuracy, thus hindering the achievement of meaningful outputs. While quantum error correction techniques \cite{Shor1995, Steane1996, Calderbank1996} could in principle offer a solution, all the proposed schemes demand technological advancements which 
are still beyond state-of-the-art capabilities. 
Within a near-term perspective~\cite{Preskill2018}, the available algorithmic implementations are therefore limited to shallow circuits applied on easy-to-prepare initial states~\cite{Peruzzo2014,Farhi2014,Wecker2015, McClean2016, Barkoutsos2018, Mazzola2019}. In most cases, only proof-of-principle experiments have been performed~\cite{O_Malley_2016,vqe1, Ganzhorn2018, extrapolate4, Ollitrault2019, Chiesa2019}, whose performances are not easily compared to classical counterparts.

Error mitigation techniques~\cite{endo2021hybrid} can in principle extend the range of operability of current technology without the need for a fully error corrected quantum device. Recently proposed solutions are based on extrapolation to zero noise~\cite{extrapolate1,extrapolate2,extrapolate3,extrapolate4, Endo2017, Barron2020}, quasi-probability distribution methods~\cite{extrapolate3, Endo2017}, symmetry enforcement~\cite{qse1, qse2, McClean2016, Sagastizabal2019, Barron2020} or stabilizer-like methods~\cite{McArdle2019}, and significant improvements have already been demonstrated experimentally~\cite{extrapolate4, Coless2018, Song2019, Sagastizabal2019}. Dedicated protocols have also been developed for mitigating measurement errors~\cite{Chen2019, Chow2010, Colm2015, extrapolate4, Nachman2019, Bravyi2020}, mostly based on the assumption that the readout noise can be simulated via a classical model where noise acts independently on each qubit.

In  this  paper,  we  introduce  a  technique  for quantum error mitigation inspired by the Lanczos algorithm for matrix diagonalization~\cite{LanczosMagic,qite}. Our method provides a rigorous way to improve ground state estimates by enhancing the projection of any given solution to the true ground state of the problem Hamiltonian, therefore representing a direct extension in the context of quantum algorithms of the well known classical Krylov space methods. These could for example be employed to increase the quality of the variational ansatz whenever the latter is limited, e.g., by the shallowness of the corresponding quantum circuit or by hardware connectivity constraints~\cite{vallury2020quantum}. In addition to that, here we go beyond a simple adaptation of classical protocols and we demonstrate a promising parallel application as a systematic quantum error mitigation procedure. The scheme remains fully compatible with the underlying variational principle and can be combined with other error mitigation techniques. Most importantly, our method does not require additional experimental resources (e.g., calibration of dedicated pulses) besides an increase in the number of observables to be measured on the quantum register. We demonstrate a practical implementation of our technique on a number of quantum physics and quantum chemistry models using  state-of-the-art quantum hardware. The general applicability of the scheme is not restricted to such problems, but can be used for any quantum computing application that requires the calculation of ground state properties.
The results indicate that the proposed error mitigation scheme can systematically provide better accuracy for energy estimates without the need for a detailed description of the underlying noise model and without being subject to specific hardware constraints.

\textit{Lanczos Algorithm.} Variational quantum algorithms are typically formulated in terms of a Hamiltonian $H$, whose smallest eigenvalue, and possibly ground eigenstate, represent the desired solution. In the variational approach, a quantum circuit with a set of tunable gate parameters $\vec{\theta}$ is used to generate an optimal approximation $|\psi(\vec{\theta}_\mathrm{opt})\rangle$ of the target quantum state in the variational subspace, using the expectation value of the Hamiltonian $\langle H \rangle_{\vec{\theta}} = \langle \psi(\vec{\theta}) | H | \psi(\vec{\theta}) \rangle$ as a cost function.  Under the effect of gate, thermalization, and readout noise, represented here as a quantum operation $\mathcal{E}$ (see also Appendix A and B and Ref.~\cite{nielson_chuang}), the variational estimate of the target ground state energy can be described in terms of a density matrix $\rho = \mathcal{E}(|\psi(\vec{\theta}_\mathrm{opt})\rangle)$  as $E=\operatorname{Tr}[\rho H]$.

Following the classical Lanczos algorithm~\cite{LanczosMagic}, both the ideal $\langle H \rangle_{\vec{\theta}}$ and the noisy energy estimates can be improved by constructing a basis of the order-$m$ Krylov subspace $\mathcal{K}^{(m)}$ through, e.g., a power iteration scheme and then diagonalizing $H$ in $\mathcal{K}^{(m)}$~\cite{LanczosMagic}. As an example, given an arbitrary pure variational state $|\psi(\vec{\theta})\rangle$, any element in $\mathcal{K}^{(2)}$ can be parametrized as $|\Psi_L(a_0,a_1,\vec{\theta})\rangle = (a_0-a_1 H) |\Psi(\vec{\theta})\rangle/\sqrt{N(a_0,a_1)}$ where $N(a_0,a_1)=\langle \Psi(\vec{\theta}) | (a_0-a_1 H)^2|\Psi(\vec{\theta})\rangle$. An improved upper bound for the true ground state energy $E_0$ can then be found by minimizing the expectation value of the Hamiltonian over the free parameters $a_0$ and $a_1$. Similarly, for any density matrix $\rho$ representing a noisy variational state, the improved energy estimate $E_{\mathrm{L}}$ can be expressed as (see Appendix~C for the general case)
\begin{equation}
E_{\mathrm{L}} = min_{a_0,a_1 \in \mathbb{R}}\frac{\mathrm{Tr}\left[ \rho H (a_0-a_1 H)^2\right]}{\mathrm{Tr}\left[\rho(a_0-a_1 H)^2 \right]}.
\label{eq:lanczos}
\end{equation}
Notice that this result coincides with the application of a single Euler step of size $\tau = a_1/a_0$ in imaginary time evolution~\cite{imag_time_1,imag_time_2,imag_time_3}. The estimate in Eq.~\eqref{eq:lanczos} can be shown to be reliable and at least as good as the initial estimate $E$ even in the presence of noise, as it still obeys the variational principle $E_0 \leq E_\mathrm{L}$ while additionally satisfying $E_\mathrm{L} \leq E$. Indeed, by defining, $\forall a_0,a_1\in\mathbb{R} $,
\begin{equation}
\rho_{\mathrm{L}}(a_0,a_1) = \frac{(a_0-a_1 H)\rho (a_0-a_1 H)}{\mathrm{Tr}\left[ \rho (a_0-a_1 H)^2 \right]},
\label{eq:rho_lanczos}
\end{equation} 
we can rewrite Eq.~\eqref{eq:lanczos} as $E_{\mathrm{L}}=\min_{a_0,a_1}\mathrm{Tr} \left[\rho_{\mathrm{L}}(a_0,a_1)H\right]$. Since $\rho_{\mathrm{L}}(a_0,a_1)$ represents a valid density matrix, i.e.,~hermiticity, semi-positive-definitness and normalization are preserved under the application of Eq.~\eqref{eq:rho_lanczos}, the variational bound $E_0 \leq E_{\mathrm{L}}$ must also hold. Moreover, the relation $E_{\mathrm{L}}\leq E$ follows from the observation that the expression minimized in Eq.~\eqref{eq:lanczos} evaluates to $E$ for $a_0=1,a_1=0$. With a similar proof, it can be shown that the same ordering $E_0 \leq E_\mathrm{L}\leq E$ is respected also for higher order ($m>2$) versions of the method.

If $\bar{\rho}_L$ is the density matrix achieving the minimum in Eq.~\eqref{eq:lanczos}, an improved estimate for any ground state observable $O$ can be obtained as $\langle  O  \rangle_\mathrm{L} = \mathrm{Tr}[\bar{\rho}_\mathrm{L} O]$. The accuracy of such estimate with respect to the true value is expected to grow, compared to the initial noisy result, as the overlap of $\rho_\mathrm{L}$ with the true ground state $|\Psi_0\rangle$ will generally be larger than the one of the original $\rho$. A sufficient condition for this to hold, if $\bar{a}_0$, $\bar{a}_1$ are the optimal parameters from Eq.~\eqref{eq:lanczos}, is that 
\begin{equation}
(\bar{a}_0-\bar{a}_1 E_\mathrm{L})^2/{\mathrm{Tr}\left[\rho(\bar{a}_0-\bar{a}_1 H)^2 \right]}>1. 
\label{eq:condition}
\end{equation}
If $\bar{a}_1\neq 0$, also $\bar{a}_0/\bar{a}_1> E_\mathrm{L}$ is required.

Intuitively, the application of the Lanczos method modifies the original spectral weights $\rho_{ii} = \langle i | \rho |i\rangle$ on the energy eigenstates $\{|i\rangle\}$, leading to the replacement
\begin{equation}
\rho_{ii} \rightarrow \bar{\rho}_{\mathrm{L},ii}  = \frac{\rho_{ii}(\bar{a}_0-\bar{a}_1 E_i)^2}{\mathrm{Tr}\left[\rho(\bar{a}_0-\bar{a}_1 H)^2 \right]}, \label{eq:spec_weight}
\end{equation}
where $E_i$ is the eigenvalue of $H$ corresponding to eigenvector $|i\rangle$. The minimum condition of Eq.~\eqref{eq:lanczos} then corresponds, similarly to the parameter shift rule in usual power iteration methods for matrix diagonalization, to a choice of $\bar{a}_0$, $\bar{a}_1$ reducing the spectral weight of excited states. 

The number of independent operators that need to be measured at order $m$ scales in the worst case as $M^{2m-1}$, where $M$ is the number of Pauli terms in the original Hamiltonian. Indeed, expectation values of the form $\langle H^n\rangle$ for $n = 0,\dots,m+1$ appear e.g.\ in Eq.~\eqref{eq:lanczos} and its generalization to $m>2$. However, through grouping of Pauli operators into tensor-product basis sets~\cite{vqe1} or by using partial state tomography and state reconstruction techniques~\cite{Huang_2020,torlaiNNtomo,PhysRevResearch.2.022060} the actual overhead in the number of measurements required by the scheme can be significantly mitigated. With extensive numerical simulations, we also find that for many relevant quantum chemistry problems encoded on $n_q$ qubits the number of Pauli strings in $H^n$ remains below the theoretical predictions ($\mathcal{O}(M^{3})$, i.e.,~$\mathcal{O}(n_q^{12})$, for $n = 3$) at least up to about $n_q = 20$, see Appendix D for details. Moreover, simple analytical arguments predict a very favorable scaling for local Hamiltonians, like for instance the ones describing spin and lattice models. In the following, we will show practical examples for which even low-order ($m=2$) results can provide estimates very close to noiseless results.

\textit{Implementation of order $m=2$ error mitigation.}
For the implementation of the method to lowest order, $E_\mathrm{L}$ is expressed in terms of the three expectation values $\langle  H^k \rangle=\mathrm{Tr}[\rho H^k]$ with $k=1,2,3$, which can all be efficiently measured in practice on a quantum processor. However, each of such estimates usually comes with an associated standard error $\sigma_{H^k}$ arising from fluctuations of the readout due to the inherent quantum statistics of measurements as well as to hardware noise. %
As a matter of fact, when such statistical errors are propagated to the final outcome $E_{\mathrm{L}}$ they can give rise to uncertainties $\sigma_{E_{\mathrm{L}}}$ which are larger than the standard errors affecting the original noisy estimate $E=\mathrm{Tr}[\rho H]$. The precision is particularly influenced by variations in time of the experimental noise levels and calibration accuracy, and can become critical at ill-conditioned values for which the minimization in Eq.~\eqref{eq:lanczos} yields a small denominator. Besides increasing the precision of the original $\langle  H^k \rangle$ estimates e.g.~via an increase in the size of the statistical samples within the same experiment, several strategies can be adopted to enhance the stability of the proposed method. A direct improvement can for example be obtained by repeating the estimation of $E_{\mathrm{L}}$ a number $n_\mathrm{repeat}$ of times, thus obtaining a set of values $\{e_{\mathrm{L},i}\}$ and of associated errors $\{\sigma_{i}\}$ which are then combined in a weighted least square (wls) average $\overline{E}_\mathrm{L,wls}$ (see also Appendix~E).
Alternatives are obtained by replacing Eq.~\eqref{eq:lanczos} with a cube root estimate $E_\mathrm{L,cube} = \sqrt[3]{\langle H^3\rangle}$, which - under some assumptions - can also enhance the spectral weight corresponding to the exact ground state (see Appendix F), or by a direct empirical choice of the parameters $a_0$ and $a_1$ in a region with low standard error yielding $E_\mathrm{L,a_0,a_1}$ (see Appendix F and G for details). It is also worth mentioning that, in principle, Eq.~(1) may become ill-conditioned whenever $|\Psi(\vec{\theta})\rangle$ is already very close to a pure eigenstate of $H$. However, in all our $m=2$ experiments we always find that that $H|\Psi\rangle$ and $|\Psi\rangle$ can be assumed to be linearly independent. In the more general case, simple numerical schemes can be introduced to detect singularities in the Krylov subspace.

\begin{figure}
\includegraphics[scale=0.94]{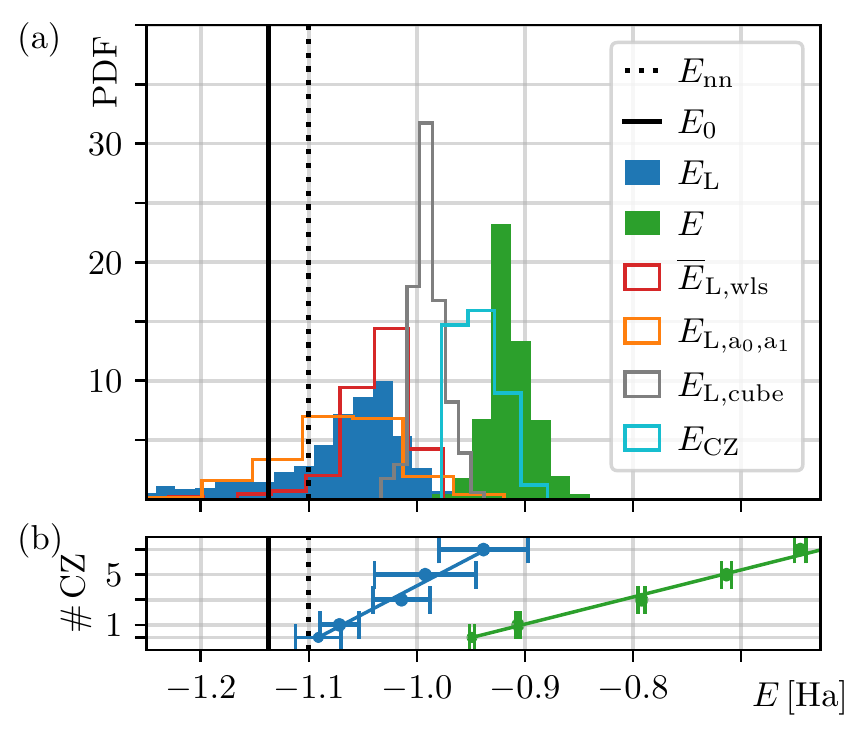}
\caption{Error mitigation for $H_2$ encoded on 4 qubits at equilibrium distance with $n_\mathrm{steps}=100$, $n_\mathrm{VQE}=10$, $n_\mathrm{shots}=8192$ run on the \textit{ibmq\_ourense} quantum processor. The ground state is approximated with the hardware efficient $R_yR_z$ variational form with one entangling layer.  (a) Probability density functions (PDF) for the bare evaluation of $\langle H \rangle_{\vec{\theta}_\mathrm{opt}}$ on real hardware (green) and the corresponding mitigated results using Eq.~\eqref{eq:lanczos}(blue), the cube root method (gray), the wls scheme (red) and the empirical selection of the $a_0$ and $a_1$ (orange). $E_0$ is the exact ground state energy, $E_\mathrm{nn}$ denotes a numerical evaluation of $\langle H \rangle_{\vec{\theta}_\mathrm{opt}}$ using the optimal VQE ansatz, whose parameters were obtained on the real quantum processor. $E_\mathrm{CZ}$ represents the value of the zero-noise extrapolation carried out on bare energy evaluations with a number of measurements comparable to the ones required for the Lanczos method. (b) Richardson extrapolation towards zero-noise energy estimations. Evaluations of $\langle H \rangle_{\vec{\theta}_\mathrm{opt}}$ are obtained by replacing every physical controlled phase (CZ) gate in the variational circuit with an odd ($1,3,5,7$) number of copies, thus systematically increasing the impact of hardware errors. Markers at $\# \mathrm{CZ} = 0$ represent the outcomes of a linear extrapolation to zero noise. The green curve is constructed with the bare energy evaluations while the data in blue are obtained by applying to each point our proposed error mitigation technique defined in Eq.~\eqref{eq:lanczos}. Combined with Richardson extrapolation, the latter procedure can recover noiseless results (dotted line) almost perfectly. The color code and the meaning of the vertical lines are consistent with the definitions in panel (a).}
\label{fig:distribution}
\end{figure}

\textit{Applications.}
In all the examples presented below, we successfully use an order $m = 2$ method to correct for hardware noise and to enhance the quality of the ans\"atze generated by the Variational Quantum Eigensolver (VQE) algorithm~\cite{vqe1,vqe2}. We use the so called $R_yR_z$ variational form (see Appendix H)~\cite{vqe1} and optimize the rotation angles $\vec{\theta}$ via the simultaneous perturbation stochastic approximation method (SPSA)~\cite{vqe1,SPSA}. To increase numerical stability and in view of the often rough optimization landscapes under study, in all cases we sample $n_\mathrm{init}=5$ parameter sets $\vec{\theta}$ from uniform distributions on $[0,2\pi]$ and use the one with the lowest energy expectation value as initial set for the optimization routine with a maximum of $n_\mathrm{steps}$ SPSA iterations. The optimization procedure is repeated a number $n_\mathrm{VQE}$ of times, choosing the $\{\vec{\theta}_\mathrm{opt}\}$ with the lowest energy as the optimal VQE solution. Error mitigation is then applied to the optimal circuit by taking measurements of the operators $H$, $H^2$ and $H^3$ and employing the methods introduced above. In particular, we obtain the direct solution of Eq.~\eqref{eq:lanczos} by analytic diagonalization of a $2\times 2$ representation of the Hamiltonian constructed from estimates of $\langle H^k\rangle$, see Appendix~C. The outcomes are compared to the original noisy estimate of the ground state energy $E$.

In the first example, we study the dissociation profile of the Hydrogen molecule $\mathrm{H}_2$~\cite{vqe1, O_Malley_2016} as a function of the internuclear distance $d$. We compute the Hamiltonian parameters using PySCF~\cite{pyscf} in the STO-3G basis set~\cite{sto3g} with the restricted Hartree-Fock method~\cite{rhf_and_uhf}, leading to a four-qubit encoding. For comparison to previous state-of-the-art work~\cite{Ganzhorn2018}, we also simulate $\mathrm{H}_2$ on two qubits exploiting the spin conservation symmetry to reduce the size of the required quantum register~\cite{taper}. 

In the second example, we increase the complexity of the simulation by focusing on the $\mathrm{H}_3$ molecule, a prototypical example of a chemical system acquiring a Berry phase under a transformation between obtuse and acute triangular shapes~\cite{h3_explain,berry_sodium_trimer,berry_phase_explain}. Here, we tune the height $h$ of an originally equilateral molecule at a constant basis side length of $\SI{1}{\angstrom}$: under this deformation, the spectrum exhibits a systematic change due to a level crossing at an arrangement of equidistant nuclei. The Hamiltonian is constructed similarly to the case of $\mathrm{H}_2$ above, but using the unrestricted Hartree-Fock method. 

As a last example, we consider the anti-ferromagnetic Heisenberg model for spin $s = 1/2$ on a tetrahedral cell
\begin{equation}
    H_\mathrm{Heis} = J\sum_{\langle i,j\rangle} \sum_{\alpha = x,y,z} \sigma_\alpha^{(i)} \sigma_\alpha^{(j)},
\end{equation}
where $J>0$, ${\sigma_\alpha}$ are the usual Pauli matrices and $\langle i,j\rangle$ denote nearest neighbors. We study the evolution of the ground state energy manifold as a function of the coupling strength of one of the bonds ($J'$) with respect to all the others. At the fully symmetric tetrahedron configuration, achieved for $J = J'$, a level crossing occurs.

\begin{figure}
\centering
\includegraphics[scale=1]{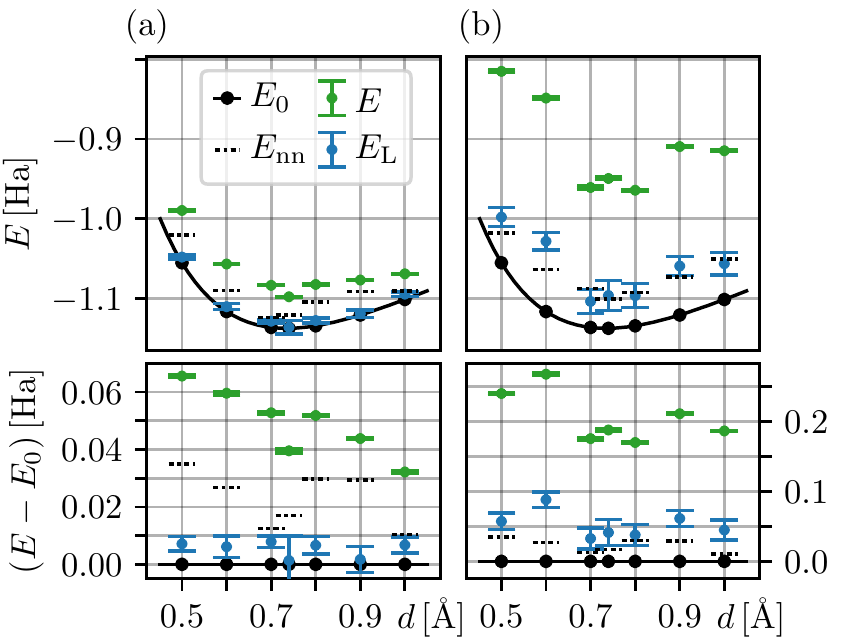}
\caption{Dissociation profiles for $\text{H}_2$ computed with 2-qubit (a) and 4-qubit (b) on \textit{ibmq\_ourense} with $n_\mathrm{steps}=100$, $n_\mathrm{VQE}=10$. Bare results (green) are compared to error-mitigated ones (blue, $n_\mathrm{shots} = 8\cdot10^5$). Black curves and dots denote the true ground state energies, and the dotted lines reference to the non-noisy energy evaluation obtained using the VQE optimal parameters resulting from (noisy) quantum hardware calculations. In all cases, the $R_yR_z$ ansatz with one entangling layer is used.}
\label{fig:h2}
\end{figure}

\textit{Results.} 
In Fig.~\ref{fig:distribution}a we report, for the 4-qubit $\mathrm{H}_2$ setup at the equilibrium bond length ($\SI{0.74}{\AA}$), the distribution of outcomes for the evaluation of the ground state energy on the \textit{ibmq\_ourense} quantum processor, accessed via the IBM Quantum Experience using the Qiskit software stack~\cite{qiskit2019}. 
The histograms are constructed by repeating 1335 times the experimental procedure, namely
the evaluation of Hamiltonian expectation values for the optimal VQE parameters $\{\vec{\theta}_\mathrm{opt}\}$ and the application of the algorithmic error mitigation method. For each data point, the noisy estimates of $\langle H^k \rangle$ were constructed by using $n_\mathrm{shots} = 8192$ runs of the relevant quantum circuit. Compared to the exact result $E_0\approx\SI{-1.137}{Ha}$, obtained with numerical diagonalization, the bare measurement of $E=\langle H \rangle_{\vec{\theta}_\mathrm{opt}}$ on the optimized ansatz (green) yields the lowest accuracy, with an average of about $\SI{-0.9145(6)}{Ha}$. A direct application of the order $m = 2$ error mitigation procedure, according to Eq.~\eqref{eq:lanczos}, significantly shifts the distribution towards the reference value: however, due to the normalization required for the Lanczos-inspired algorithm, the distribution (blue) has a long tail towards small energies and large uncertainty. By increasing $n_\mathrm{shots}$ one can directly obtain smaller uncertainties $\sigma_{H^k}$ on the evaluations of $\langle H^k \rangle_{\vec{\theta}}$ within a single experiment. Despite being computationally expensive (typically $\sigma_{H^k}\propto 1/\sqrt{n_\mathrm{shots}}$), this approach leads in our case to an average mitigated result $E_\mathrm{L}=\SI{-1.084(7)}{Ha}$ with $n_\mathrm{shots} = 10^7$. Less demanding alternatives are the application of the cube root method (grey), yielding the smallest improvement on the energy evaluation but the most stable results, the weighted least square over groups of $n_\mathrm{repeat}=5$ repetitions (red), or the empirical selection of the $a_0$ and $a_1$ parameters (orange). The latter procedure is carried out by choosing $a_0$ and $a_1$ in such a way to keep the average estimated measurement uncertainty below $\SI{0.02}{Ha}$, which is far smaller than the investigated energy scales, while still providing an improvement over the bare results.
In fact, by tuning the ratio $a_0/a_1$ one can interpolate between the outcome of Eq.~\eqref{eq:lanczos} and a more stable result, ultimately converging to the original bare value for $a_0/a_1 \rightarrow \infty$ (see also Appendix G). It is worth mentioning that a reasonable scale for the energy uncertainty, and therefore suitable values for $a_0$ and $a_1$, can be identified through a systematic procedure involving only noisy energy estimates (see Appendix~I) without prior knowledge of ideal results.

\begin{figure}
\centering
\includegraphics[scale=0.96]{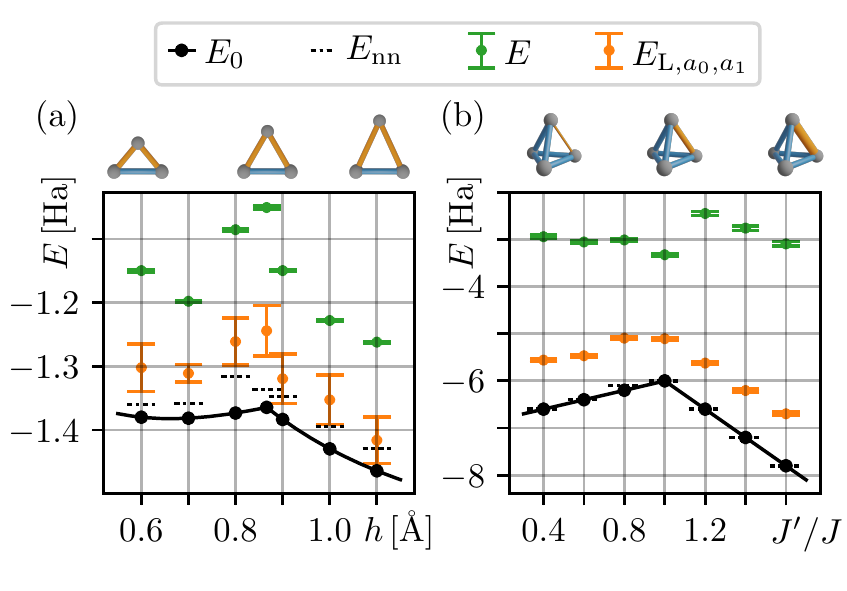}
\caption{Ground state energies for the triangular $\text{H}_3$ molecule (a) and for the Heisenberg tetrahedron model (b, $J'$ corresponds to the orange bond). Bare results (green) are compared to error-mitigated ones (orange) obtained by fixing $a_0$ and $a_1$ to achieve a maximal estimated standard error of $\SI{0.04}{Ha}$. Black curves and dots denote the true ground state energies, dotted lines are VQE results optimized numerically and evaluated in the absence of noise. The $R_yR_z$ ansatz with 1 (3) entangling layers is used, with $n_\mathrm{steps}=2000$, $n_\mathrm{VQE}=4$~$(10)$, $n_\mathrm{shots} = 8\cdot 10^5$~$(8192)$, for $\text{H}_3$ (respectively the Heisenberg model).}
\label{fig:tet_h3}
\end{figure}

The performance of the proposed methods in presence of high noise values is assessed in Fig.~\ref{fig:distribution}b. Here, for the same configuration of Fig.~\ref{fig:distribution}a, we artificially increase the noise levels by replacing every controlled phase (CZ) operation in the entangling block of the variational circuit with an odd ($3$, $5$, $7$) number of copies~\cite{Tacchino2019,Stamatopoulos2020}. 
We observe a dominantly linear dependence on the number of CZ, with an increase of the uncertainties associated to the mitigated results $E_\mathrm{L}$ for higher levels of noise. Fig.~\ref{fig:distribution}b also shows the application of the Richardson zero-noise extrapolation (ZNE) procedure~\cite{extrapolate1,extrapolate4} on both bare energy measurements and Lanczos-mitigated ones, reporting an almost complete cancellation of the hardware noise when both error mitigation strategies are combined. 

To compare the Lanczos and the bare ZNE methods on an equal footing, we also take into account the overall number of measurements required in the two cases. In Fig.~\ref{fig:distribution}a, we report an additional histogram, $E_\mathrm{CZ}$, representing the distribution of error-mitigated energy reconstructions obtained with the zero-noise extrapolation (ZNE) procedure applied to the bare energies, using a similar amount of resources (i.e., measurements) as needed for the proposed Lanczos-inspired procedure (i.e., $E_L$ results). 
For the latter, each of the data points contributing to the histogram is obtained by measuring $15$ Pauli strings for the expectation value of the Hamiltonian $\langle H\rangle$ and $32$ Pauli strings for $\langle H^2\rangle$ and $\langle H^3\rangle$. 
In total, this amounts to $79$ Pauli strings measured independently using $8192$ shots each. 
To match as close as possible these overall requirements, we increase the statistics for the energy expectation reconstructions with the ZNE approach. 
More specifically, for each ZNE realization we measure four times $\langle H\rangle$, one for each choice of $\#\mathrm{CZ}=1,3,5,7$. 
This consists of a total of $4\cdot 15 = 60$ Pauli strings. For each Pauli string we double the number of shots used ($2\cdot8192$) to estimate the expectation values, such that 
%In this way, 
the overall number of measurements is actually in favour of the ZNE method. 
The whole procedure is repeated $50$ times to construct the corresponding histogram. 
These results show that the use of more measurements for the ZNE method (keeping fixed the range of $\#\mathrm{CZ}$) only slightly reduces the uncertainties, without significantly improving the average energy expectation value (see also Fig.~\ref{fig:probing_different_zne_shemes} in Appendix~N). 
In fact, we consistently observe that the Lanczos method yields results closer to the true ground state energy, although with typically larger uncertainties. 
It is also worth stressing that we do not in general consider (as shown in Fig.~\ref{fig:distribution}b) the Lanczos and ZNE methods as alternative options, but rather as complementary approaches.

The results for the full dissociation profile of $\mathrm{H}_2$ encoded on 2 and 4 qubits are shown in Fig.~\ref{fig:h2}. The proposed error mitigation scheme enhances the quality of the energy estimates up to a factor of 5. In particular, for the 2-qubit calculation this leads to an accuracy of about $\SI{10}{mHa}$, which is competitive with state-of-the-art results obtained, e.g., on dedicated hardware~\cite{extrapolate1,Ganzhorn2018} and comparable to the application of quantum subspace expansion (QSE) in Ref.~\cite{Coless2018}. 
We notice that for $m=2$ the Lanczos algorithm requires a smaller subspace size than QSE at the cost of a larger number of measurements (i.e., more Pauli strings).
It is also worth pointing out that in some cases our Lanczos-inspired error mitigation provides corrections going beyond a mere noise-cancellation effect in energy evaluations, and which can be interpreted as an improvement on the ansatz itself. This can be seen by comparing the mitigated results with what would be obtained form an ideal noiseless measurement $E_\mathrm{nn}$, simulated numerically starting from the variational ansatz learned on the real hardware (dashed lines). Additional quantitative details on both the improvement over noiseless energy estimates, due to the very nature of Krylov-subpace methods, and the fulfillment of Eq.~\eqref{eq:condition},
are provided in the Appendices~J and~K)

Finally, in Fig.~~\ref{fig:tet_h3} we report the outcomes for the $\mathrm{H}_3$ dissociation curve and for the Heisenberg tetrahedron model mapped on 6 and 4 qubits, respectively. Due to the significant increase in complexity and size of the required hardware simulations, for these models the optimal VQE parameters $\vec{\theta}_\mathrm{opt}$ were computed via numerical simulations, while the final energy evaluations and the corresponding mitigated results were obtained by implementing the optimal circuit on the \textit{ibmq\_almaden} ($H_3$) and \textit{ibmq\_ourense} (tetrahedron) quantum processors. At difference with Fig.~\ref{fig:h2}, here we made use of the empirical selection method for $a_0$ and $a_1$ in order to reduce the standard error on the mitigated points $E_{L,a_0,a_1}$ (see Appendices for additional data). In both cases, the Lanczos-inspired scheme significantly helps in recovering the most relevant ground state features, i.e., the extremal points of the energy surfaces and level crossings.

\textit{Summary.}
We introduced a hardware agnostic technique to mitigate errors on near term quantum processors and we described its direct application to quantum chemistry and quantum physics problems. The proposed approach significantly improves the calculation of ground state properties by effectively enhancing the spectral weight of noisy variational ansatzes on the respective optimal solutions while ensuring the physical consistency of the outcomes. This method does not require direct control on low-level hardware operations and can in principle be combined with other available noise mitigation protocols. We reported experimental results for energy evaluations closely approaching the exact reference values and demonstrated the robustness of the approach in several test cases with highly non-trivial ground state properties. The computational cost of this error mitigation procedure is ultimately associated to an increase in the number of measurements to be performed and can potentially be reduced by the introduction of more elaborate schemes for the grouping of Pauli terms appearing in the target Hamiltonian. In  view  of  these  arguments, we conclude that the proposed method has the potential to become a standard error mitigation scheme for any type of noisy quantum computing platform.

\textit{Acknowledgements.} 
Authors acknowledge interesting discussions with Igor O. Sokolov. 
This project has received funding from the European Research Council (ERC) under the European Union’s Horizon 2020 research and innovation programm (ERC-StG-Neupert-757867-PARATOP). 
IT acknowledges the financial support from the Swiss National Science Foundation (SNF) through the grant No. 200021-179312. IBM, the IBM logo, and ibm.com are trademarks of International Business Machines Corp., registered in many jurisdictions worldwide. Other product and service names might be trademarks of IBM or other companies. The current list of IBM trademarks is available at \url{https://www.ibm.com/legal/copytrade}.

\bibliographystyle{unsrtnat}
\bibliography{main}

\newpage

\appendix

\onecolumngrid

\section{Kraus Operator Expression for Errors} \label{sec:readout_errors}

The validity of the Lanczos method requires a density matrix description of the state of the quantum computer with noise. This is guaranteed as long as the whole evolution of the state of the quantum computer can be described with Kraus operators. Indeed, in the basic noise model of qiskit, the noise of the quantum computer is divided into gate errors, thermalization errors, and readout errors. While the Kraus operator expression for the gate and thermalization errors are well known and given below, the Kraus operator expression for the readout error is derived in  section~\ref{sec:kraus_readout}. With each application of a gate an error occurs with a probability $p_{\mathrm{depol.\,error}}$. These errors are modeled via the depolarization channel, described for one qubit as
\begin{equation}\begin{split}
    \rho_{\mathrm{n,depol.}} &= (1-\gamma_1) \rho_0 + \gamma_1 \mathds{1}/{2^N} = \sum_i K_i \rho_0 K_i^\dagger,\\
  K_0 &= \sqrt{1-3\gamma_1/4}\, \mathds{1},\\
  K_1 &= \sqrt{\gamma_1/4}\, \sigma_x, \\
  K_2 &= \sqrt{\gamma_1/4}\, \sigma_y, \\
  K_3 &= \sqrt{\gamma_1/4}\, \sigma_z,\\
\gamma_1 &=p_{\mathrm{depol.\,error}},
\end{split}
\end{equation}
returning the density matrix $\rho_{\mathrm{n,depol.}}$ including the gate error given the state of the quantum computer after the application of the respective gate $\rho_0$~\cite{nielson_chuang}.

The thermalization error occurring during a gate duration of $t$ is modelled using two different error sources. These are the phase flip and the generalized amplitude damping channel with respective Kraus operator expressions

\begin{equation}\begin{split}
\rho_{\mathrm{n,p. flip}} &= \sum_i K_i \rho_0 K_i^\dagger,\\
K_0 &= \sqrt{1-p_f/2} \, \mathds{1},  K_1 = \sqrt{p_f/2} \, \sigma_z,\\
p_f &= 1-e^{-t/\tau_1},
\end{split}
\end{equation}
where $\tau_1$ is a decay time to be fitted, and

\begin{equation}\begin{split}
\rho_{\mathrm{n,gad}} &= \sum_i K_i \rho_0 K_i^\dagger,\\
K_0 &= \sqrt{p} \begin{pmatrix} 1 & 0 \\ 0 & \sqrt{1-\gamma_2} \end{pmatrix},  K_1 = \sqrt{p} \begin{pmatrix} 0 & \sqrt{\gamma_2} \\ 0 & 0 \end{pmatrix},\\  K_2 &= \sqrt{1-p} \begin{pmatrix} \sqrt{1-\gamma_2} & 0 \\ 0 & 1 \end{pmatrix},  K_3 = \sqrt{1-p} \begin{pmatrix} 0 & 0 \\ \sqrt{\gamma_2} & 0 \end{pmatrix},\\
p &= \left( e^{-\omega\over{k_bT}}+1\right)^{-1},\gamma_2=1-e^{-t/\tau_2},
\end{split}
\end{equation}
where $\omega$ is the energy difference between the two qubit levels, $T$ the temperature and $\tau_2$ a decay time. While the phase flip channel describes a decay of the off-diagonal elements of a one-qubit density matrix, with the generalized amplitude damping channel a decay  of the diagonal elements towards the Fermi-Dirac distribution is achieved.

For each applied gate these Kraus expressions are applied consecutively on the involved qubits using the noise calibration data of an IBM quantum device to find $\tau_1,\tau_2,p_\mathrm{depol. \, error},t,T$ and $\omega$.

\section{Kraus operator expression for readout errors}\label{sec:kraus_readout}

In this section we explain the simulation of readout errors in qiskit and derive a Kraus operator expression so as to incorporate the readout errors by changing the density matrix.

 We consider an operator $O$ in the Pauli basis
\begin{align*}
O = \sum_{ \bm{i} } o^{\bm i} \sigma_{i_0} \otimes \sigma_{i_1} \otimes\dots \otimes \sigma_{i_{N-1}}
\end{align*}
with an expectation value of
\begin{align}
\langle O \rangle = \sum_{ \bm{i} } o^{\bm i} \langle \sigma_{i_0} \otimes \sigma_{i_1} \cdots \rangle = \sum_{ \bm{i},\bm s = \{s_0,s_1,\ldots,s_{N-1}\} } o^{\bm i} p^{\bm i}(\bm s) \underbrace{\prod_j s_j}_{=s},
\end{align}
where $s^{j}\in\{-1,1\}$ is the measurement outcome of $\sigma_{i_j}\in \{\sigma_0=\mathds{1},\sigma_x,\sigma_y,\sigma_z \}$ with a measurement probability of $p^{\bm i}(\bm s)$ for the measurement result $\bm s\in\{-1,1\}^{\otimes N}$ for the Pauli string $\bm i$. Now, this probability changes with the readout error  $p_r^{\bm i}(\bm s|\bm s')$ for a given Pauli string $\bm i$, i.e., it encodes the probability to measure $\bm s$ instead of $\bm{s'}$ given that the measurement outcomes of the single qubits were $\bm {s'}=\{s'_{0},\ldots\}$. Now, the measurement result with readout error (r.e.) can be written as
\begin{align}
\langle O \rangle_\mathrm{r.e.} & = \sum_{ \bm{i},\bm s } o^{\bm i} p_\mathrm{r.e.}^{\bm i}(\bm s)s ,\\
p_\mathrm{r.e.}^{\bm i}(\bm s) &= \sum_{\bm s'} p_r^{\bm i}(\bm s|\bm s')p^{\bm i}(\bm s').
\end{align}
In qiskit, $p_r^{\bm i}(\bm s|\bm s')$ is described using the probability for a single qubit flip $p_{r,j}^{i_j}(s_j|s'_j)$ on qubit $j$, which is an approximation making the probability that a readout error occurs on one qubit independent of the measurement result of all other qubits. Furthermore, we assume that $p_{r,j}^{i_j}(1|-1)=p_{r,j}^{i_j}(-1|1)$ implying $p_{r,j}^{i_j}(1|1)=p_{r,j}^{i_j}(-1|-1)$. If this assumption is not fulfilled, the variational principle is not fulfilled in general anymore (see counterexample below). Intuitively, if $p_{r,j}^{i_j}(1|-1)\neq p_{r,j}^{i_j}(-1|1)$, the operator expression becomes dependent on the measurement result, so that a Kraus operator expression cannot be constructed anymore.
This yields
\begin{align}
p_r^{\bm i}(\bm s|\bm s') = \prod_j p_{r,j}^{i_j}(s_j|s'_j) 
\end{align}
so that 
\begin{align}
\langle O \rangle_\mathrm{r.e.} & = \sum_{ \bm{i},\bm s ,\bm s'} o^{\bm i}  \left[\prod_j s_j p_{r,j}^{i_j}(s_j|s'_j)\right] p^{\bm i}(\bm s')= \sum_{ \bm{i} ,\bm s'} o^{\bm i}  \left[\prod_j p_{r,j}^{i_j}(1|s'_j)-p_{r,j}^{i_j}(-1|s'_j)\right] p^{\bm i}(\bm s').
\end{align}
This allows us to further simplify
\begin{align}
p_{r,j}^{i_j}(1|s'_j)-p_{r,j}^{i_j}(-1|s'_j)=s'_j [p_{r,j}^{i_j}(s'_j|s'_j)-p_{r,j}^{i_j}(-s'_j|s'_j)]=s'_j [1-2p_{r,j}^{i_j}(-s'_j|s'_j)].
\end{align}
To emphasize the independence of $p_{r,j}^{i_j}(-s'_j|s'_j)$ on $s'_j$ by assumption we write $p_{r,j}^{i_j}\equiv p_{r,j}^{i_j}(-s'_j|s'_j)$. This allows us to write 
\begin{align}
 \langle O \rangle_\mathrm{r.e.}= \sum_{\bm i,\bm s'}  o^{\bm i} \left[\prod_j (1-2 p_{r,j}^{i_j})\right] s' p^{\bm i}(\bm s')=\sum_{\bm i}  o^{\bm i} \left\langle\bigotimes_j \sigma_{i_j} (1-2 p_{r,j}^{i_j})\right\rangle.
\end{align}

To be able to write down Kraus operators, we further assume that $p^{x}_{r,j}=p^{y}_{r,j}=p^{z}_{r,j}$ which we refer to by $p_{r,j}$ in the following. This assumption is reasonable, as the only difference in the measurement of $\sigma_x,\sigma_y,\sigma_z$ is one single qubit rotation, which has a small error, which is neglected here. Furthermore, we have $p^{\mathds{1}}_{r,j}=0$ as the identity does not need to be measured. 
Then, we can write 
\begin{align}
\sigma_{i_j}(1-2p_{r,j}^{i_j})=\epsilon_j(\sigma_{i_j})
\end{align}
with 
\begin{align}
\epsilon_j(\sigma_{i_j})&=  \sigma_{i_j}(1-2p_{r,j}) +p_{r,j}(\sigma_{i_j} + \sigma_{x,j} \sigma_{i_j} \sigma_{x,j} + \sigma_{y,j}  \sigma_{i_j}\sigma_{y,j} + \sigma_{z,j} \sigma_{i_j} \sigma_{z,j})/2\\
&=  \sigma_{i_j}(1-2p_{r,j}) +p_{r,j} \mathrm{Tr}(\sigma_{i_j})\mathds{1}.
\end{align}
Using this expression, it becomes evident, that $\epsilon_j$ is a Kraus operator. This simplifies the error calculation to 

\begin{align}
 \langle O \rangle_\mathrm{r.e.}=\sum_{\bm i}  o^{\bm i} \left\langle\bigotimes_j \epsilon_j(\sigma_{i_j})\right\rangle=\sum_{\bm i}  o^{\bm i} \left\langle \epsilon\left( \bigotimes_j \sigma_{i_j}\right)\right\rangle=\langle \epsilon(O)\rangle,
\end{align}
with $\epsilon=\epsilon_0 \circ \epsilon_1\cdots$ and $\epsilon_j$ acts trivially on all sites except for site $j$.

Finally, we can use that
\begin{align*}
\langle O \rangle_{\mathrm{r.e.}} = \mathrm{Tr}(\rho \epsilon(O)) = \mathrm{Tr}(\epsilon(\rho) O)
\end{align*}
to describe the readout error by changing the density matrix.

\,

\,

\paragraph*{Violation of the Variational Principle for $p_{r,j}^{i_j}(1|-1)\neq p_{r,j}^{i_j}(-1|1)$}
The variational principle is not in general fulfilled anymore for $p_{r,j}^{i_j}(1|-1)\neq p_{r,j}^{i_j}(-1|1)$. For this, consider the example $H=- \sigma_x \otimes \sigma_x -  \sigma_y \otimes \sigma_y- \sigma_z \otimes \sigma_z=-2({\bm S}^2-2{\bm s}^2)=-2(S(S+1)-2s(s+1))$ with spectrum $\{-1,-1,-1,3\}$. One eigenstate with eigenvalue $-1$ is $\Psi_{S=0,S_z=1}= |00\rangle$.
Assuming equal readout error probabilities for the two qubits $p_{r,j}^{x,y,z}(s|-s)=p(s|-s)$ the expectation value of the Hamiltonian becomes
\begin{align}
    \langle 00 | H | 00 \rangle = -1+4p(1|0)(1-p(1|0))-2(p(0|1)-p(1|0))^2,
\end{align}
which can be below the minimal eigenvalue of $-1$ for $p(1|0)\neq p(0|1)$ and as shown larger equals $-1$ for $p(1|-1)=p(-1|1)$.

\section{Relation to the Lanczos Algorithm and Generalization}\label{sec:rel_to_lanczos}
In the Lanczos algorithm, the ground state is approximated using the ground state of the Hamiltonian expressed in an order $m$ Krylov space $\mathcal{K}^{(m)}$. The matrix expression for the Hamiltonian is real and symmetric, as all matrix entries are real functions of $\{\langle \Psi | H^l | \Psi\rangle \in \mathds{R}\}_{l=0}^{2m-1}$. Thus, every eigenstate can be chosen to be real. The ground state is the linear combination with real coefficient with minimal eigenvalues of all elements of the Krylov space. Since the minimal eigenvalue is the lowest possible measurement outcome for all states in $\mathcal{K}^{(2)}$ the minimization in Eq.~(1) cannot yield a lower energy estimate. Furthermore, the linear combination of states parametrized with $a_0$ and $a_1$ spans $\mathcal{K}^{(2)}$ so that the ground state lies within the space parametrized with $a_0$ and $a_1$. Hence, the minimization in Eq.~(1) yields the ground state of the Hamiltonian expressed in $\mathcal{K}^{(2)}$ and Eq.~(1) evaluates to its eigenvalue.

This holds also for higher orders $m$ for the generalized expression of Eq.~(1) that reads
\begin{equation}
    E_{\mathrm{L},k,n,m} = \min_{\bm a \in \mathbb{R}^m} \sqrt[k]{\frac{\langle \Psi | H^k (\sum_{i=0}^{m-1} a_i H^i)^n |\Psi \rangle}{\langle \Psi | (\sum_{i=0}^{m-1} a_i H^i)^n |\Psi \rangle}}, \label{eq:general}
\end{equation}
for which $E_{\mathrm{L},k,n,m}>E_0$ holds with odd $k$ and even $n$. This follows by generalization of the statements of the main text, this section and Appendix~\ref{sec:appendix_cube_root}. For $k=1,n=2$ expression~\eqref{eq:general} is equivalent to the Lanczos method for the smallest eigenvalue in an order-$m$ Krylov space, as can be shown with the same argument as for $\mathcal{K}^{(2)}$. Furthermore, $E_{\mathrm{L},k=1,n,m}\leq E $, as $E$ is the special case of $a_0=1,a_{i>0}=0$. \\

The direct relationship with the classical Lanczos algorithm immediately suggests a practical way to solve Eq.~\eqref{eq:lanczos}. For the 2-dimensional case ($m=2$) presented in this work, the minimization is in fact equivalent to the diagonalization of a $2\times 2$ matrix, which can be done analytically. The matrix is constructed by expressing all matrix elements in terms of the measurement results for $\langle H \rangle_{\vec{\theta}},\langle H^2 \rangle_{\vec{\theta}}$ and $\langle H^3 \rangle_{\vec{\theta}}$. The matrix elements can be obtained by constructing $\mathcal{K}^{(2)}=\{|v_1\rangle = |\Psi(\vec{\theta})\rangle,|v_2\rangle= (H-\alpha_1)|\Psi(\vec{\theta})\rangle/\beta_2\}$ with $\alpha_1 = \langle H \rangle_{\vec{\theta}}, \beta_2 = \sqrt{\langle H^2 \rangle_{\vec{\theta}}-\langle H \rangle_{\vec{\theta}}^2}$. 
In this basis, the matrix expression for the Hamiltonian reads
\begin{align}
    H_{\mathcal{K}^{(2)}} = \begin{pmatrix} \langle v_1 | H | v_1\rangle & \langle v_1 | H | v_2\rangle \\\langle v_2 | H | v_1\rangle &\langle v_2 | H | v_2\rangle \end{pmatrix} =  \begin{pmatrix} \alpha_1 & \beta_2 \\ \beta_2 & \alpha_2 \end{pmatrix}
\end{align}
where $\alpha_2 =(\langle H^3 \rangle_{\vec{\theta}}-2\langle H^2 \rangle_{\vec{\theta}}\langle H \rangle_{\vec{\theta}}+\langle H \rangle_{\vec{\theta}}^3)/\beta_2^2$. 
In higher orders (i.e., $m > 2$), the minimization can in principle be performed in a similar way by constructing and diagonalizing, in a classical post-processing stage, the corresponding matrix $H_{\mathcal{K}^{(m)}}$. 
Alternatively, one could also achieve the desired result with an iterative minimization procedure whose cost function, parametrized by $a_0$, $a_1$, is defined in Eq.~\eqref{eq:lanczos} of the main text. 
A similar approach is also possible at higher orders, when a suitable parametrization to navigate the appropriate Krylov subspace is considered.\\

Concerning the regularization techniques proposed in the main text and discussed in more details in some Appendix sections below, they can in principle be generalized straightforwardly to higher orders $m$.
This is for instance true for the direct application of the Lanczos method (with the generalized version of Eq.~\eqref{eq:lanczos} presented in Eq.~\eqref{eq:general} above) as well as for the closely related weighted least squares averaged version. 
The same Eq.~\eqref{eq:general} shows how the empirical tuning of the parameters $a_i$ could scale for higher orders, with $m-1$ independent parameters to optimize for an order $m$ mitigation. 
Although it is true that such approach could become impractical for large values of $m$, a viable solution could be to apply an iterative minimization procedure of a suitably parametrized cost function, which can be constructed from the noisy estimates of powers of the Hamiltonian. 
This could possibly be used to explore the neighborhood of the formal solution obtained with, e.g., the numerical diagonalization of the reduced Hamiltonian in the Krylov subspace. 
Finally, the cube root version can be replaced with similar higher order constructions of the form $\sqrt[k]{\langle H^k\rangle_{\vec{\theta}}}$, for odd $k$.

\section{Scaling}

In general terms and in the absence of more specific assumptions, when two-body interactions are present $H^2$ and $H^3$ are, respectively, 8th and 12th order operators, with correspondingly large numbers of Pauli strings to be measured in order to reconstruct average values. However, as we argue below through an extensive collection of numerical investigations, the reality for many cases of practical interest shows effective results which are much more favorable.

\begin{figure}
    \centering
    \includegraphics[width=0.45\textwidth]{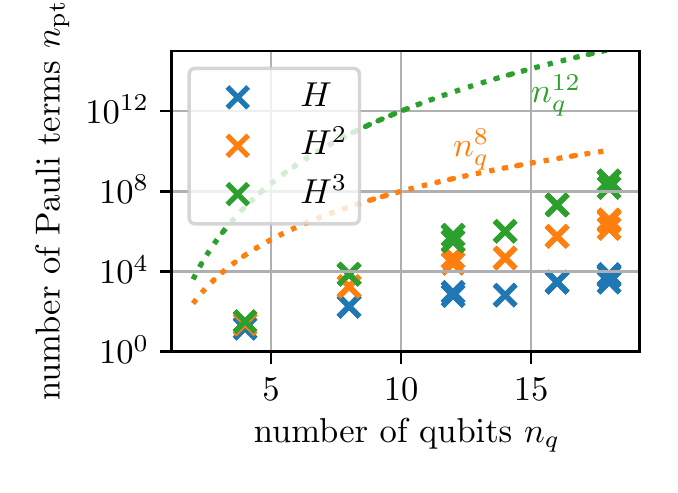}
    ~
    \includegraphics[width=0.45\textwidth]{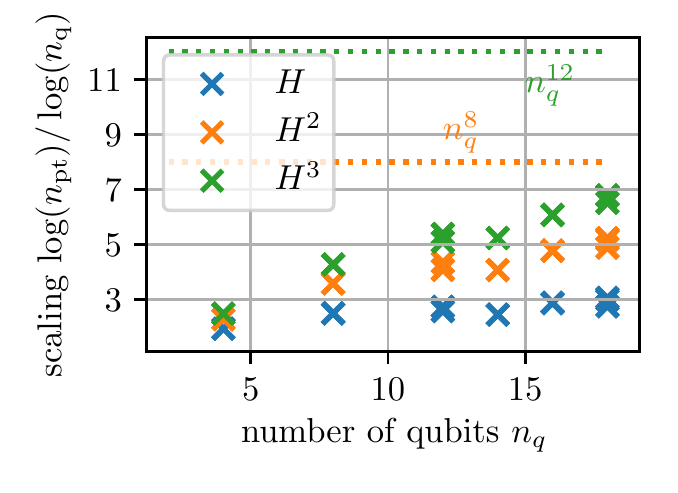}
    \caption{Number of Pauli terms $n_\mathrm{pt}$ for different real molecules encoded on $n_\mathrm{q}$ qubits. In the left panel the values are shown using a logarithmic scale, while in the right panel the exponent $y$ is shown, so that $n_\mathrm{q}^y=n_\mathrm{pt}$. The terms of the simulated molecules are calculated in the sto3g basis if not stated otherwise: H$_2$ ($n_q=4$), H$_2$ ($n_q=8$, basis: 631g), H$_4$ ($n_q=8$), H$_6$ ($n_q=12$), LiH ($n_q=12$), H$_2$O ($n_q=14$), NH$_3$ ($n_q=16$), H$_8$ ($n_q=16$), H$_{10}$ ($n_q=18$), NN ($n_q=18$) and NaH ($n_q=18$), where the $18$-qubit molecules are simulated using a two-qubit reduction scheme for implementation purposes. The Pauli terms for the Hamiltonian $H$ are calculated using PySCF and the H$_2$ equilibrium interatomic distance of $\SI{0.75}{\AA}$ and $\SI{1.6}{\AA}$ for LiH. The geometry of the molecules was chosen to be one dimensional except NH$_3$, which was simulated on a tetrahedron. The results for similar number of qubits $n_q$ tend to agree for different molecules.}
    \label{fig:scaling_lanczos_numerical}
\end{figure}

Focusing on quantum chemistry problems, in Fig.~\ref{fig:scaling_lanczos_numerical}, we report the number of Pauli terms corresponding to the Hamiltonian operators $H^n$ ($n = 1,2,3$) for a collection of 11 different molecules which require up to $n_q=18$ encoding qubits. These were obtained numerically without resorting to grouping techniques, which might provide a further reduction of approximately an order of magnitude in the number of independent measurement settings required for the reconstruction of average values. We observe, in particular, that the number of terms in $H$ seems to saturate around a power law of the form $n_q^3$, making it effectively a 3rd-order operator. This behaviour is typical whenever conservation laws are present, which provide implicit constraints to the number of available physical interactions. Regarding the terms $H^2$ and $H^3$, although the trend is actually still far from achieving a power law behavior with a well established exponent, we can already explicitly observe that the actual number of Pauli strings remains far from the predicted values $\mathcal{O}(n_q^{8})$ for $H^2$, respectively $\mathcal{O}(n_q^{12})$ for $H^3$, and even below the $\mathcal{O}(n_q^{6})$ (respectively $\mathcal{O}(n_q^{9})$) that could be inferred by the 3rd order nature of the $H$ operators employed. We interpret such results by mentioning two concurrent effects: on one hand, the commutation algebra of the Pauli strings effectively reduces the number of terms appearing in successive products of $H$. On the other hand, the total number of independent Pauli strings that can be constructed with $n_q$ qubits equals $4^{n_q}$, and can be quickly saturated for small and medium sized systems. While we cannot exclude that in the large $n_q$ limit an actual 8th or 12th order scaling is followed, the onset of such regime appears to be still beyond many of the examples of current practical interest in near term applications. The latter represent, in fact, some ideal candidates for our proposed error mitigation strategy.

Much more favourable scaling properties can also be predicted, independently of the size of the problem, for local Hamiltonians: There, the number of terms scales as $n_q$, so that the Lanczos method scales in the worst case like $n_q^3$. This class of Hamiltonians is represented in the main text by the Heisenberg tetrahedron. Local Hamiltonians constitute by themselves an interesting field of research with many connections, e.g., to the theory of magnetism (see also a recent work by Vallury et al.~\cite{vallury2020quantum}, where the authors analyse an application of Krylov methods for local Hamiltonians), and include among others many examples of spin and lattice systems.

Finally, it is worth noticing that all the analysis conducted in the present work remains valid whenever a different observable $O$ is used instead of the Hamiltonian. In such case, only $OHO$ has to be measured, while the variational principle is still fulfilled and the ground state energy estimate will not be worse than the noisy prediction. In fact, $H$ can actually be used to guide the choice of the operator $O$, which can be as small as the quantum computer is capable of dealing with. However, an analysis of this strategy requires a more detailed study which is left for future work.

\section{$\mathrm{H}_2$ Saturation of wls}

\begin{figure}
\centering
\includegraphics[width=0.5\textwidth]{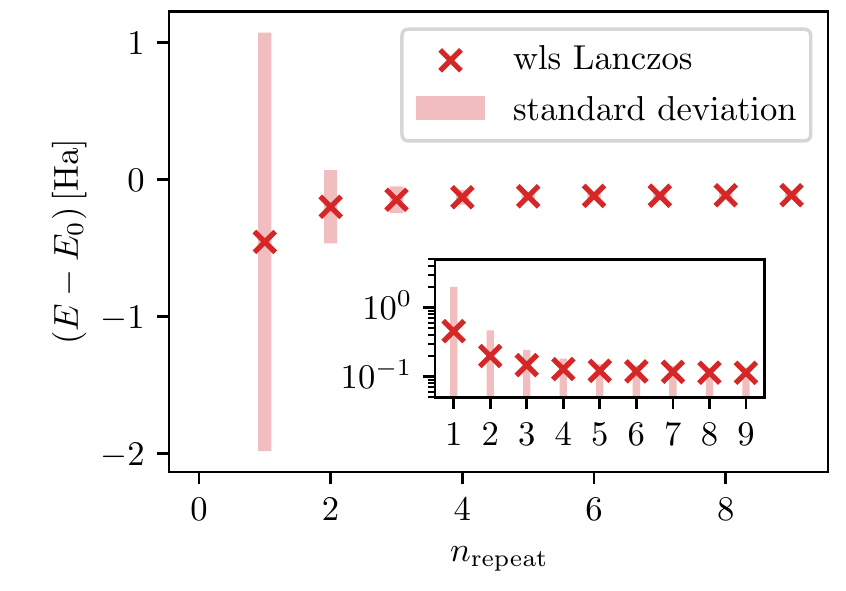}
\caption{The results of the dependence of the wls method (red) on the number of Lanczos estimates ($n_\mathrm{repeat}$) used is shown for a distance of $\SI{0.74}{\AA}$ for the simulation of $\mathrm{H}_2$ on four qubits.
The results are obtained by using the $100$ measurements on the quantum computer and choose $n_\mathrm{repeat}$ of them randomly $1000$ times. The mean of the results and the standard deviation of the results is shown for the wls. 
As a conclusion for $n_\mathrm{repeat} \gtrsim 5$ we obtain similar results, validating the use of $n_\mathrm{repeat}=5$ in the main text. 
}
\label{fig:h2_wls_convergence}
\end{figure}

In this section we show the results of taking the weighted least square average (wls) over $n_\mathrm{repeat}$ measurement outcomes. 
In this case the error mitigated energy estimate can be obtained via the formula:
\begin{equation}
    \overline{E}_\mathrm{L,wls} = \frac{\sum_{i=0}^{n_\mathrm{repeat}-1} e_{\mathrm{L},i}\sigma_{i}^{-2}} {\sum_{i=0}^{n_\mathrm{repeat}-1} \sigma_{i}^{-2}},
\end{equation} with standard error $\sigma_{\overline{E}_\mathrm{L},\mathrm{wls}}^{-2} = \sum_i \sigma_{i}^{-2}$.
Notice that this solution fulfills the generalized Gauss-Markov Theorem~\cite{gen_mtheorem}, and therefore $\overline{E}_\mathrm{L,wls}$ has minimal variance.
The results are shown in Fig.~\ref{fig:h2_wls_convergence}, where we plotted the dependency of the mean of the wls result and its standard deviation against varying $n_\mathrm{repeat}$. As expected, for increasing $n_\mathrm{repeat}$ the standard deviation of the wls decreases faster than $1/\sqrt{n_\mathrm{repeat}}$, which is validating the generalized Gauss-Markov Theorem. Furthermore, we can observe an increase of the mean of the wls result with increasing $n_\mathrm{repeat}$. The reason is the small uncertainty for results with larger energy, which therefore have a larger weight. The higher $n_\mathrm{repeat}$, the more unlikely are wls results which only contain low energy estimates with larger measurement uncertainty. Hence, the mean of the wls result increases. The mean for $n_\mathrm{repeat}=1$ is below the true ground state energy showing that this value might not be useful and reflecting the necessity of considering the estimated measurement uncertainties.

\section{\label{sec:appendix_cube_root}Variational Principle for the Cube Root Method}
For the cube root method the variational principle holds as well since for any odd $k$
\begin{equation}
    \langle H^k \rangle = \sum_i \alpha_i E_i^k \geq \sum_i \alpha_i E_0^k = E_0^k
\label{eq:cuberootgen}
\end{equation}
and $\sqrt[k]{.}$ is monotone restricting the image of $\sqrt[k]{.}$ to the real axis. 

The effective enhancement of the ground state spectral weight can be understood by comparing the expectation value
\begin{align}
    \langle H^k\rangle_{\vec{\theta}} = \sum_i (\alpha_i E_i^{k-1}) E_i \label{eq:new_method}
\end{align}
with 
 \begin{align}
     \langle H \rangle_{\vec{\theta}}^k = \sum_i (\alpha_i \langle H \rangle_{\vec{\theta}}^{k-1}) E_i, \label{eq:old_method}
 \end{align}
where $\alpha_i$ represents the probability of observing the $i$-th energy eigenstate on a given variational (noisy) ground state estimate. If we assume that the dominant contributions come from the true ground state ($i=0$) and other eigenstates for which $|E_i|<|E_0|$, i.e.,
\begin{align}
    \sum_{0,i:|E_i|<|E_0|} \alpha_i \gg \sum_{i:|E_i|\geq |E_0|} \alpha_i,
\end{align}
we have that the weight associated to the ground state energy in the expression of Eq.~\eqref{eq:new_method} is enhanced by a factor $E_0^{k-1}/\langle H \rangle_{\vec{\theta}}^{k-1} > E_i^{k-1}/\langle H \rangle_{\vec{\theta}}^{k-1}$, which is larger for the ground state, with respect to the quantity in Eq.~\eqref{eq:old_method}. 
This assumption is reasonable, for example, in those cases where the ground state energy and the energies of the lowest excited states are negative. 
Although the cube root method is not \textit{a priori} guaranteed to be helpful in all cases, the fact that it complies with the variational principle makes it straightforward to discard the result $\sqrt[k]{\langle H^k\rangle_{\vec{\theta}}}$ whenever this is larger than the unmitigated value $\langle H \rangle_{\vec{\theta}}$. 
In such case, one could infer that the (noisy) variational quantum state features a significant occupation of high energy eigenstates.

\section{\label{sec:generalisation2}{Dependence of the Lanczos Energy Estimate on $a_0,a_1$}}

\begin{figure}
\centering
\includegraphics[scale=1]{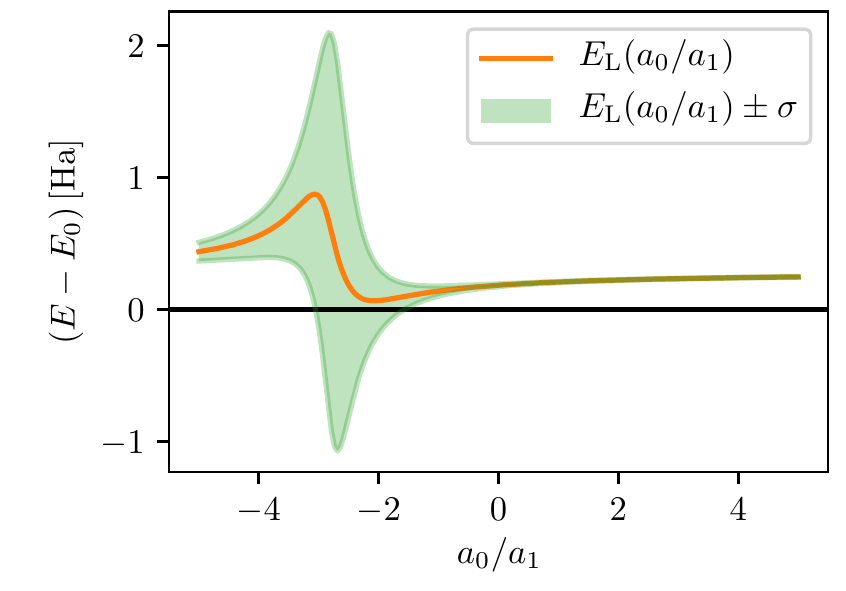}
\caption{ The dependence of the mean of $100$ energy estimates using formula (1) 
for fixed $a_0/a_1$ with reference to $E_0$ and the mean of the estimated standard errors of the $100$ results are shown for an height of $\SI{0.8}{\angstrom}$ for $\text{H}_3$. 
In general, we observe that for larger $a_0/a_1$ the standard deviation decreases, while the energy estimate increases. For $a_0/a_1\approx E_0$ the standard deviation is the largest as most of the weight of the statevector is projected out, the norm is close to $0$. In conclusion, when fixing the parameter $a_0/a_1$ a value needs to be chosen which provides a good compromise between increased ground state energy estimates and standard errors. For $a_0/a_1 \rightarrow \infty$ the method converges to the bare measurement of $\langle H \rangle_{\vec{\theta}_\mathrm{opt}}$. In order to reduce the standard error by a factor of $1/\sqrt{n}$, $n$ estimates can be averaged over. 
}
\label{fig:h3_variation_d_clean}
\end{figure}

\begin{figure}
\centering
\includegraphics[scale=1]{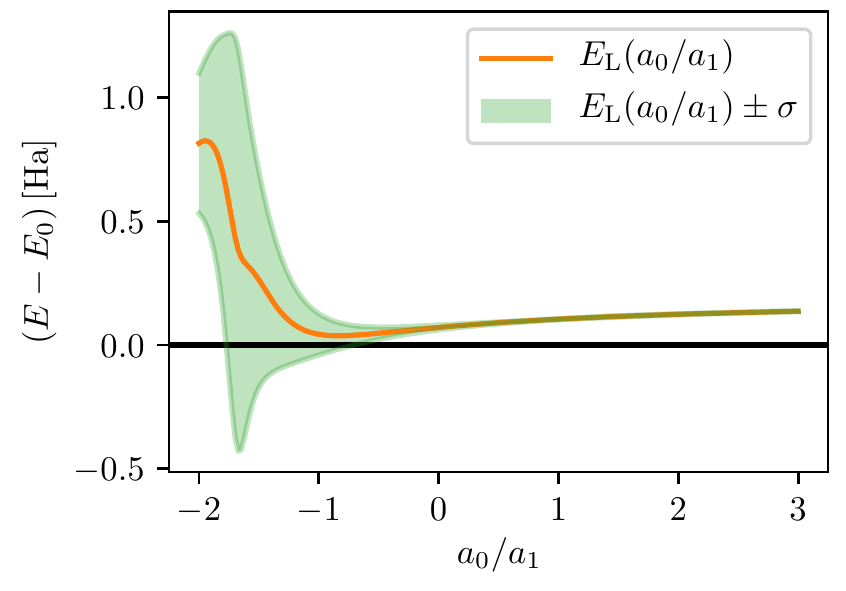}
\caption{The dependence of the mean of $100$ energy estimates using formula (1) %
for fixed $a_0/a_1$ with reference to $E_0$ and mean of the estimated standard errors of the $100$ results are shown for an internuclear distance of $\SI{0.74}{\angstrom}$ for $\text{H}_2$ on four qubits.
In general, we observe that for larger $a_0/a_1$ the standard deviation decreases, while the energy estimate increases. For $a_0/a_1\approx E_0$ the standard deviation is the largest as most of the weight of the statevector is projected out, the norm is close to $0$. In conclusion, when fixing the parameter $a_0/a_1$ a value needs to be chosen which provides a good compromise between increased ground state energy estimates and standard errors. For $a_0/a_1 \rightarrow \infty$ the method converges to the bare measurement of $\langle H \rangle_{\vec{\theta}_\mathrm{opt}}$. In order to reduce the standard error by a factor of $1/\sqrt{n}$, $n$ estimates can be averaged over. 
}
\label{fig:h2_variation_d_clean}
\end{figure}

In Fig.~\ref{fig:h3_variation_d_clean} and \ref{fig:h2_variation_d_clean} the dependence of the energy estimate on the choice of parameters $a_0,a_1$ is shown. For $\mathrm{H}_3$ even for large number of measurements we could observe an instability of the method. This is related to the fact, that the measurements have been taken over a long time period of six days during which the noise fluctuates significantly. This instability can be detected considering the estimates for the measurement uncertainty. This is shown in Fig.~\ref{fig:h3_variation_d_clean}, where the measurement uncertainty also increases for too small energy estimates. In all cases considered, the ground state energy was found to be within a likely number of standard deviations of the measurement result so that the measurement uncertainty reflects the accuracy of the method sufficiently well.

\newpage

\section{Variational Form}\label{si:var_form}
\begin{figure}
\centering
\begin{center}
\qquad{\small
\Qcircuit @C=0.5em @R=.6em {
\lstick{|q_{0} \rangle}  & \gate{R_yR_z} & \ctrl{1} & \qw      & \qw      & \qw      & \gate{R_yR_z} & \qw & \qw \\
\lstick{|q_{1} \rangle}  & \gate{R_yR_z} & \gate{Z} & \ctrl{1} & \qw      & \qw      & \gate{R_yR_z} & \qw & \qw \\
\lstick{}                &               &          &          &          &          &               &     &     \\
\lstick{\vdots}          &               &          &          &          &          &               &     &     \\
\lstick{}                &               &          &          &          &          &               &     &     \\
\lstick{|q_{N-2}\rangle} & \gate{R_yR_z} & \qw      & \qw      & \gate{Z} & \ctrl{1} & \gate{R_yR_z} & \qw & \qw \\
\lstick{|q_{N-1}\rangle} & \gate{R_yR_z} & \qw      & \qw      & \qw      & \gate{Z} & \gate{R_yR_z} & \qw & \qw \gategroup{1}{7}{7}{3}{1.1em}{--} \\
}
}
\end{center}
\caption{This figure depicts an example of an $R_yR_z$ circuit.}
\label{fig:var_form}
\end{figure}
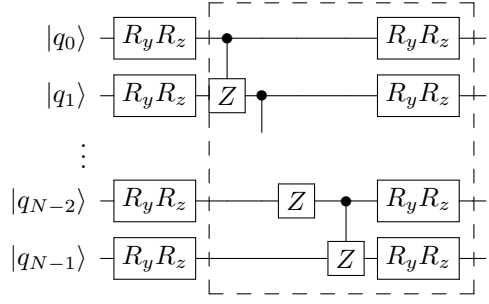

In this section the used variational form, namely the $R_yR_z$ ansatz, is shown in Fig.~\ref{fig:var_form}. It consists of a layer of initial rotations followed by entanglement layers. One such entanglement layer is shown below denoted with the box. This entanglement layer is repeated as often as required to obtain the highest achievable accuracy.

\section{Empirical selection of the parameters $a_0,a_1$}

\begin{figure}
    \centering
    \includegraphics[scale=0.85]{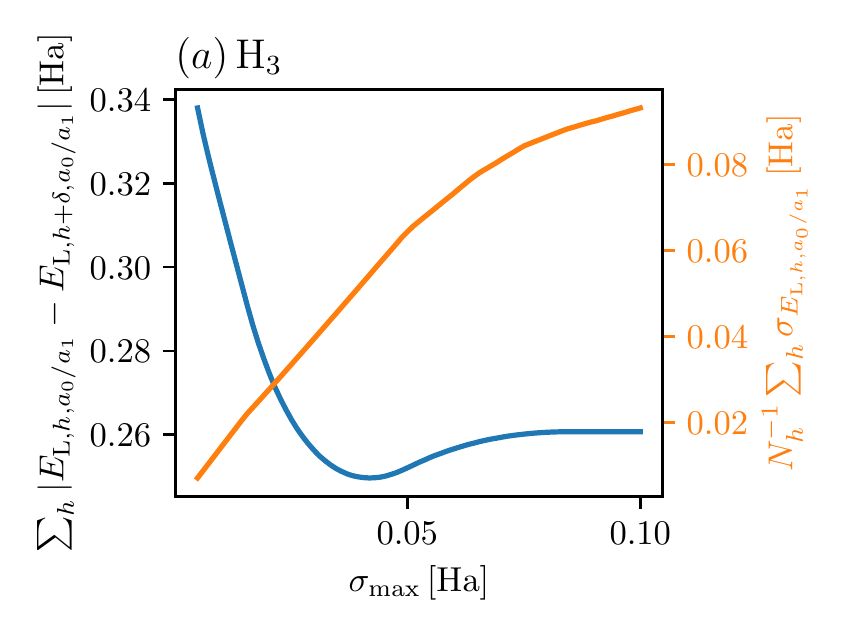}
    \includegraphics[scale=0.85]{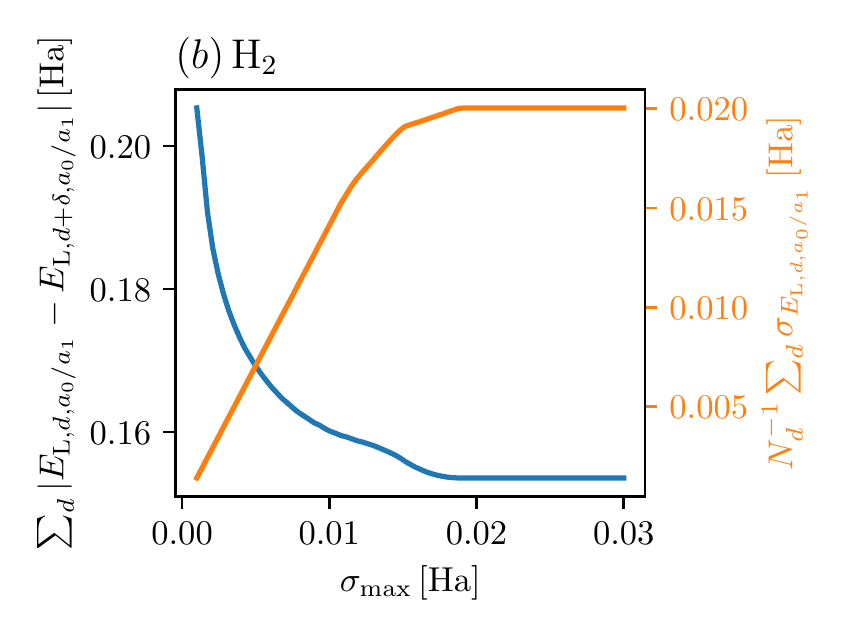}
    \caption{The dependence of the summed level distance of consecutive parameters is shown in (a) for $\mathrm{H}_3$ and (b) for $\mathrm{H}_2$ for different chosen cutoffs $\sigma_\mathrm{max}$, which is used for all heights $h$ (distances $d$ for $\mathrm{H}_2$). Given the cutoff and the system parameter $h$, $a_0/a_1$ is chosen as minimizing the ground state energy estimate yielding $E_{\mathrm{L},h}$  (Eq.~(1) main text) if $\sigma_{E_{\mathrm{L},h}}<\sigma_\mathrm{max}$ or as the smallest value $a_0/a_1$ so that $\sigma_{E_\mathrm{L},h,a_0/a_1}<\sigma_\mathrm{max}$. We observe a saturation for large $\sigma_\mathrm{max}$, since all estimated errors are below the cutoff so that it is not applied. For small cutoffs, the $\Delta_h E$ increases up to the non-error mitigated value. In orange the estimated noise error on $E_{\mathrm{L},h,a_0/a_1}$ averaged over $h$ is shown. Its saturation for large cut offs $\sigma_\mathrm{max}$ reflects the circumstance that the chosen cutoff is larger than the estimated standard error $\sigma_{E_\mathrm{L}}$.}
    \label{fig:sum_spacing}
\end{figure}

In this chapter we want to give a quantitative description for how the bound of $\sigma_\mathrm{max}=\SI{40}{\,\mathrm{mHa}}$ %
on the energy estimates $E_{\mathrm{L},h,a_0/a_1}$, where $h$ is the varied system parameter, is calculated. Such value was chosen by comparing the results for the whole dissociation profiles for different values of a maximal measurement uncertainty $\sigma_\mathrm{max}$, essentially identifying reasonable energy scales from typical energy differences in the system obtained from noisy estimates. This corresponds to a modification of Eq.~(1) in the main text to $E_{\mathrm{L},a_0,a_1} = \min_{a_0,a_1|C} \mathrm{Tr}(\rho_{\mathrm{L}}(a_0,a_1) H)$ with the additional constraint $C:\sigma_{E_{\mathrm{L},a_0,a_1}}\leq\sigma_\mathrm{max}$, where $\sigma_{E_{\mathrm{L},a_0,a_1}}$ is the estimated measurement uncertainty on $E_{\mathrm{L},a_0,a_1}$.
Note that reasonable values of $\sigma_\mathrm{max}$ can be found only in the range $[\sigma_{E},\sigma_{E_{\mathrm{L},h}}]$. To quantify the choice of $\sigma_\mathrm{max}$, we evaluate the sum of the energy level distances for consecutive ground state energy estimates
\begin{align}
\Delta_h E = \sum_h |E_{L,h,a_0/a_1}- E_{L,h+\delta,a_0/a_1}|,
\end{align}
which we expect to be the lowest for a more pronounced energy profile (i.e.~by requiring adjacent energy estimates to be close and therefore reducing noise induced fluctuations, this strategy highlights peaks or discontinuities in the profile, if they are present, without actually making any assumptions on the existence of such features). This is shown in Fig.~\ref{fig:sum_spacing} (a) for $H_3$ and (b) for $H_2$, for the tetrahedron $\sigma_{E_{\mathrm{L},J'/J}}<\sigma_\mathrm{max}$, for any $\sigma_\mathrm{max} \lessapprox |E_{L,J'/J,a_0/a_1}- E_{L,J'/J+\delta,a_0/a_1}|$ so that $\Delta_{J'/J} E$ is constant. We denote with $\sigma_{E_{\mathrm{L},J'/J}}$ the measurment uncertainty for the energy estimate of Eq.~(1) in the main text.

We stress that choosing a $\sigma_\mathrm{max}$ is always legitimate to do, as the results are bounded from below by the true values via the variational principle and from above by the non-mitigated, noisy results. Hence, by tuning the results as a whole an improvement within these boundaries is expected.

In summary, the proposed error mitigation procedure is particularly well suited to refine energy profiles and to bring to light extremal features in the spectrum. Towards this goal, we suggest to choose a common bound on the standard deviation in such a way that the calculated energy differences are the smallest. This does not involve any requirement regarding knowledge of the solution a priori.

\section{Groundstate overlap improvement}

In the main text a sufficient condition for the improvement of the overlap between the true ground state and the ground state estimate is presented. It requires the fulfillment of Eq.~(3) of the main text and ${a}_0/{a}_1>E_L$ for the optimal ${a}_0/{a}_1$ for $a_1\neq 0$. In Tab.~\ref{tab:equation_3_results} we report the results for the evaluation of such condition for all the data points presented in the main text. We recall that whenever the presented value is larger than one, an improvement of the overlap between the ground state estimate and the true ground state is expected.
For each of the values provided, we have that ${a}_0/{a}_1>E_L$ and, as reported in the Table, we evaluate the left hand side of Eq.~(3)
\begin{equation}
(\bar{a}_0-\bar{a}_1 E_\mathrm{L})^2/{\mathrm{Tr}\left[\rho(\bar{a}_0-\bar{a}_1 H)^2 \right]} 
\end{equation}
to be larger than $1$ for all cases. Eq.~$(3)$ is therefore fulfilled and the overlap between the ground state and the estimate is improved.

\begin{table}[h]
    \centering
    \resizebox{\columnwidth}{!}{%
    \begin{tabular}{l|l|l|l|l|l|l|l|c}
    \begin{tabular}{c}parameter\\system\end{tabular} & \multicolumn{6}{c}{values with uncertainties} & &\begin{tabular}{c}mean,\\standard deviation\end{tabular}\\ \hline
$d \, [$\AA$]$ & $0.50$ & $0.60$ & $0.70$ & $0.74$ & $0.80$ & $0.90$ & $1.00$ & \\
$\mathrm{H}_2$ $2$-qubit & $1.039(4)$ & $1.042(7)$ & $1.033(4)$ & $1.04(2)$ & $1.039(6)$ & $1.05(1)$ & $1.025(6)$ & $1.038 ,\, 0.007$ \\ \hline
$d \, [$\AA$]$ & $0.50$ & $0.60$ & $0.70$ & $0.74$ & $0.80$ & $0.90$ & $1.00$ & \\
$\mathrm{H}_2$ $4$-qubit & $1.15(2)$ & $1.15(2)$ & $1.14(3)$ & $1.16(4)$ & $1.14(4)$ & $1.17(3)$ & $1.18(4)$ & $1.16,\, 0.01$ \\ \hline
$h \, [$\AA$]$ & $0.60$ & $0.70$ & $0.80$ & $0.87$ & $0.90$ & $1.00$ & $1.10$ & \\
$\mathrm{H}_3$ & $1.11(4)$ & $1.09(2)$ & $1.15(5)$ & $1.15 (4)$ & $1.15(5)$ & $1.11(5)$ & $1.17(6)$ & $1.13,\, 0.03$ \\ \hline
$J'/J$ & $0.60$ & $0.70$ & $0.80$ & $0.87$ & $0.90$ & $1.00$ & $1.10$ & \\
Heisenberg & $1.341(9)$ & $1.310(9)$ & $1.261(8)$ & $1.216 (7)$ & $1.341(9)$ & $1.342(8)$ & $1.324(8)$ & $1.31,\, 0.05$ \\
    \end{tabular} }
    \caption{The results for the evaluation of the left side of Eq.~$3$ are reported for the four considered systems in the main text: $\mathrm{H}_2$ simulated on $2$ and $4$ qubits, $\mathrm{H}_3$ and the Heisenberg spin chain. The values are given for each parameter configuration used for the respective system with the uncertainty on the last digit due to the measurement uncertainty on $H,H^2$ and $H^3$. In the last column, the average of all the values for each system is given as well as the standard deviation of the values around the given mean.}
    \label{tab:equation_3_results}
\end{table}

\section{Noiseless Results Improvement}

In this section, we provide insights about the improvement of the Lanczos estimates over noiseless energy estimates. This effect is a direct consequence of the application of the Lanczos diagonalization method within Krylov subspaces. Indeed, similarly to a classical diagonalization problem, any noiseless simulation of a variational ansatz essentially corresponds to an approximation of the eigenvector of the target Hamiltonian associated to the minimum eigenvalue. It is therefore not surprising that the quality of such approximation can benefit from the application of the Lanczos method, as this is explicitly designed to increase the spectral overlap with the exact solution. We notice, for example, that this appears to be the main motivation behind some recent work by Vallury et al.~\cite{vallury2020quantum}. As we discuss in the main text, the aim of our work is, besides remarking the usefulness of this effect for quantum variational algorithms, to demonstrate its very promising parallel application as a systematic error mitigation procedure.

\begin{figure}[h!]
    \centering
    \includegraphics{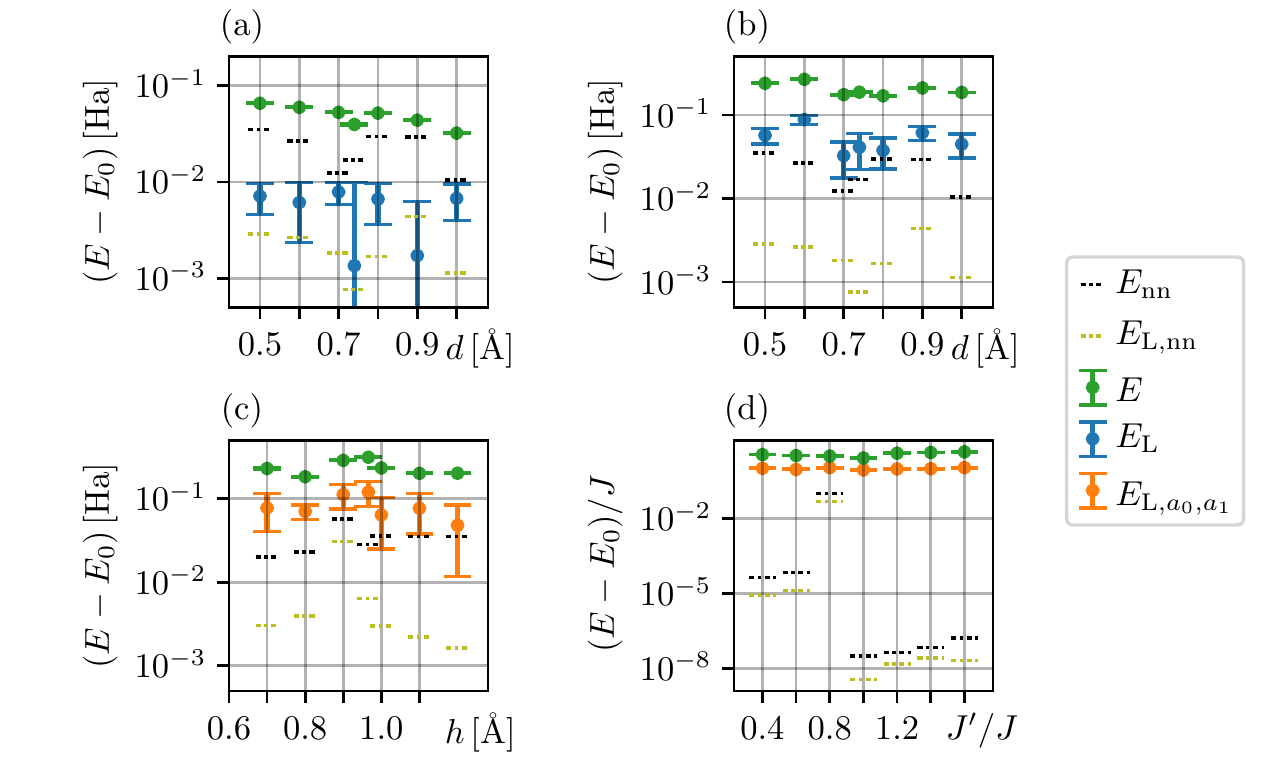}
    \caption{Here we show in addition to the data presented in Fig. 2 and Fig. 3 in the main text the improvement obtained via the application of the Lanczos algorithm on the noiseless results for the four cases discussed in the main text (a) $\mathrm{H}_2$ simulated on two qubits, (b) $\mathrm{H}_2$ simulated on four qubits, (c) $\mathrm{H}_3$ and (d) the Heisenberg model.}
    \label{fig:my_label}
\end{figure}

For completeness, we report in Fig.~\ref{fig:my_label} the results obtained from numerical noiseless simulations of the VQE algorithm for the four example models presented in the main text. These are obtained with Qiskit statevector simulator, which works out the linear algebraic calculations without the effect of sampling or hardware noise. It shows that the application of the method described in this work provides an improvement towards the true ground state values, which is purely due to the effective increased spectral overlap provided by the Lanczos procedure.

\section{Convergence Rate}

In principle, due to the delicate scaling properties of the method already at the lowest $m = 2$ order, we do not expect that higher orders will be used extensively. Hence, we do not provide a detailed numerical study for the convergence of the method as a function of $m$. Instead we can make use of the Kaniel-Paige convergence theory for density matrices. This yields a convergence $|E_0-E_L|<\mathrm{constant} \cdot r^{n}$, where $n$ is the order of the method and $r$ is a convergence rate. This corresponds to what was observed when considering examples. For more information, see e.g.\ chapter 10.1.5 in Ref.~\cite{golub2013matrix} (in combination with the trigonometric definition of the Chebyshev polynomials).

\section{Effect of shot noise and error propagation}

\begin{figure}
    \centering
    \includegraphics{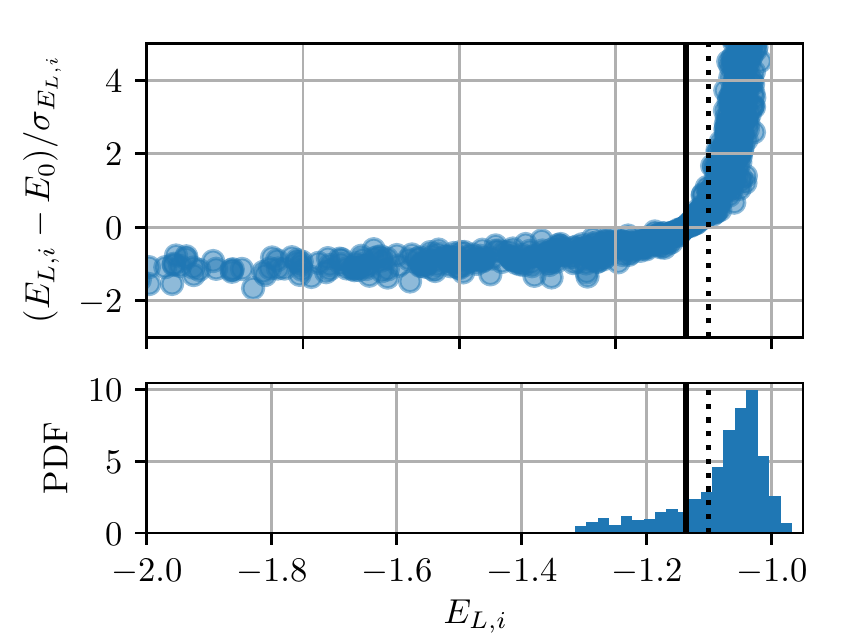}
    \caption{In this figure we show additional properties of the numerical data, which is also presented in Fig.~1 of the main text. For each data point we show on the $x$ axis the energy estimate and, on the $y$ axis, the corresponding deviation from the true ground state value in terms of the measurement uncertainty. In the lower panel the histogram of the data in the upper panel is shown, which is the same as presented in the main text in Fig. 1. Like in the figure in the main text, the solid black line marks $E_0$, the dashed black line the noiseless estimate $E_\mathrm{nn}$.}
    \label{fig:rel_noise_error}
\end{figure}

In this section, we provide additional insights about the results for $E_L$ reported in Fig.~1 of the main text, where some data points placed below the true ground state energy appear to contradict the validity of the variational principle. We support our discussion with the analysis presented in Fig.~\ref{fig:rel_noise_error}, where more detailed information is provided particularly regarding the uncertainty associated to the mitigated energy estimates. As it can be seen, the long tails of the energy distribution should be interpreted as a consequence of shot noise and error propagation effects. Indeed, as we highlighted in the main text and in Appendices~\ref{sec:readout_errors} and ~\ref{sec:kraus_readout}, all other possible sources of errors act as quantum channels, thus always yielding valid density matrices. However, statistical fluctuations due to the finite size of the samples taken from a quantum hardware can for example lead to apparent violations of the variational bounds whenever the minimization procedure in Eq.~(1) of the main text yields an ill-conditioned denominator. Nevertheless, as we report in Fig.~\ref{fig:rel_noise_error}, such extreme cases are always clearly marked by large intrinsic errors accompanying the energy estimate and can thus be clearly identified and marked as unphysical, in what essentially represents a self-consistence mechanism for the proposed method.

\section{Additional Figures}

\begin{figure}
\centering
\includegraphics[scale=0.8]{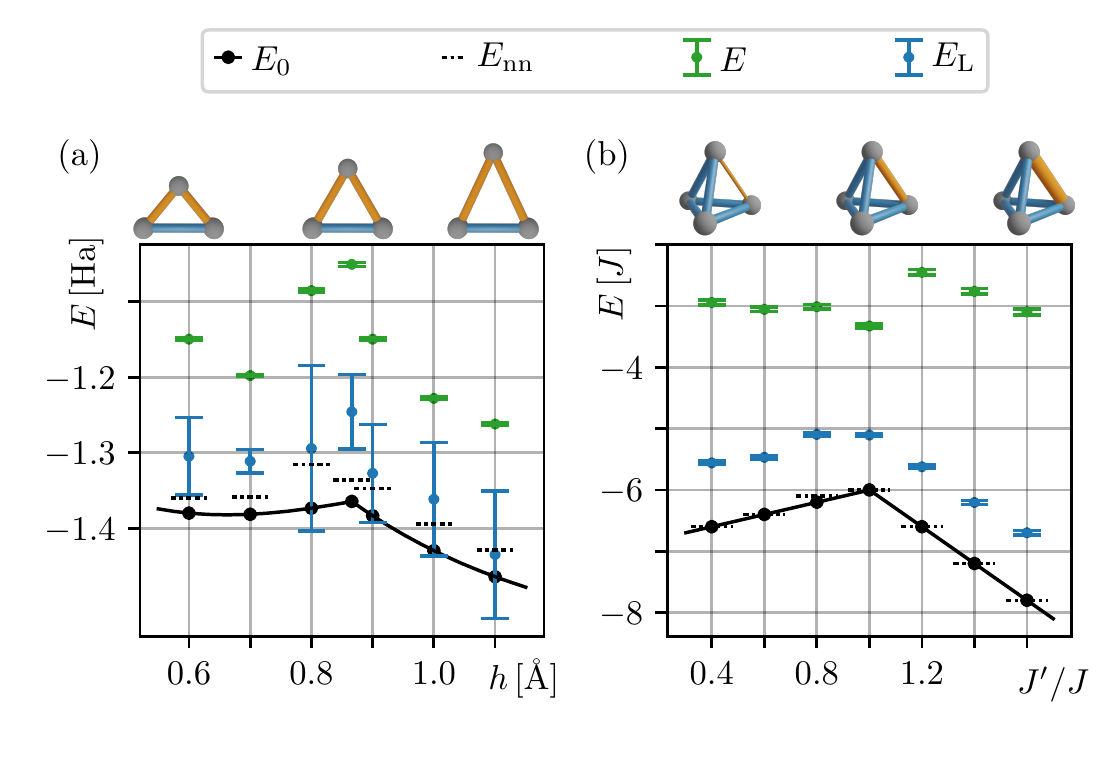}
\caption{ The results for the VQE algorithm for the Heisenberg tetrahedron and $\text{H}_3$ simulated on four qubits with ibmq\_ourense and on six qubits with {ibmq\_almaden}  in dependence of the systems parameters are depicted in a) and b), respectively. The colors denote the same results as in Fig. 1 in the main text, blue are the results of the Lanczos algorithm and the results of the bare measurement of the Hamiltonian are shown in green. Black curves and dots denote the true ground state energies, dotted lines are VQE results optimized numerically and evaluated in the absence of noise. The $R_yR_z$ ansatz with 1 (3) entangling layers is used, with $n_\mathrm{shots}=8\cdot 10^5$~$(8192)$, $n_\mathrm{steps}=2000$, $n_\mathrm{VQE}=4$~$(10)$, for $\text{H}_3$ (respectively the Heisenberg model).}
\label{fig:tet_h3_alma_lan}
\end{figure}

\begin{figure}
\centering
\includegraphics[scale=0.8]{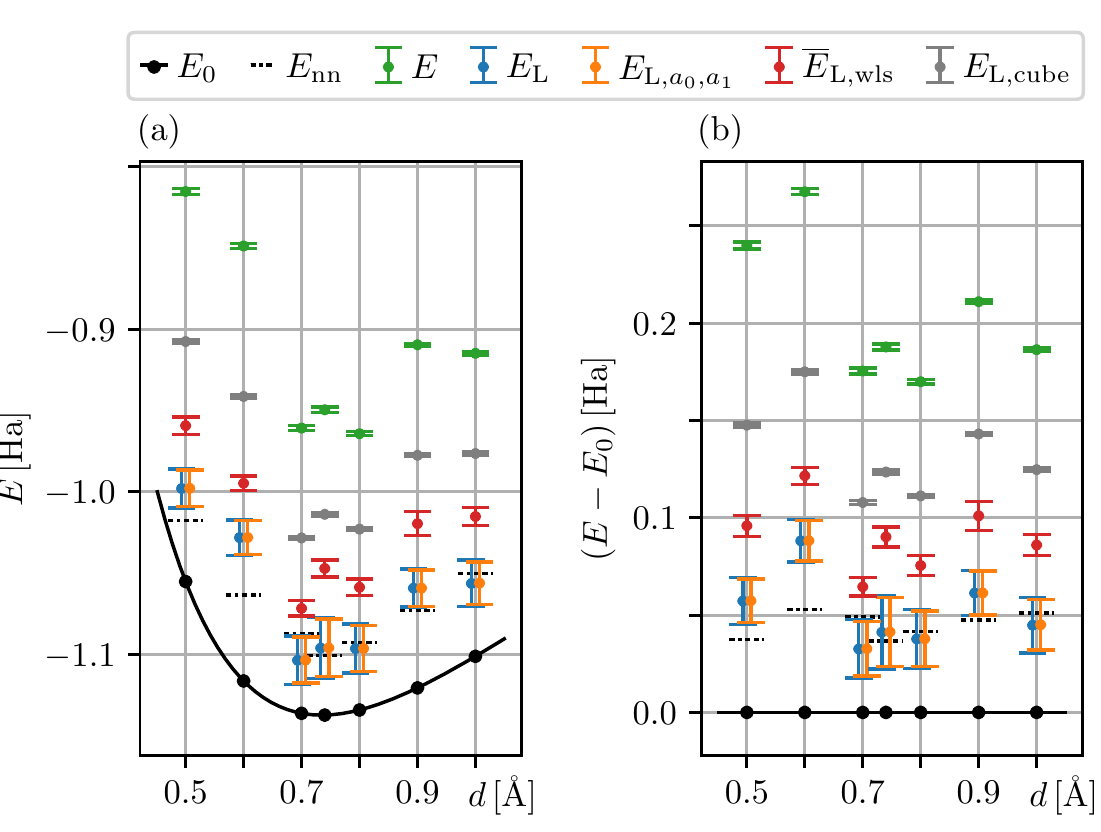}
\caption{The results for the VQE algorithm for $\text{H}_2$ simulated on four qubits in dependence of the system configuration are depicted in (a) alongside with their accuracy in (b). The colors denote the same results as in Fig. 1 in the main text, blue are the results of the Lanczos algorithm, orange are the results of the fixed parameter method, gray are the results of the cube root method, red are the averaged results of taking the wls over five measurements, while the results of the bare measurement of the Hamiltonian are shown in green. Black denotes the true ground state energy and the dotted line reference to the non-noisy energy result obtained with the circuit optimized on the (noisy) quantum computer.}
\label{fig:diss_h2_full}
\end{figure}

\begin{figure}
    \centering
    \includegraphics{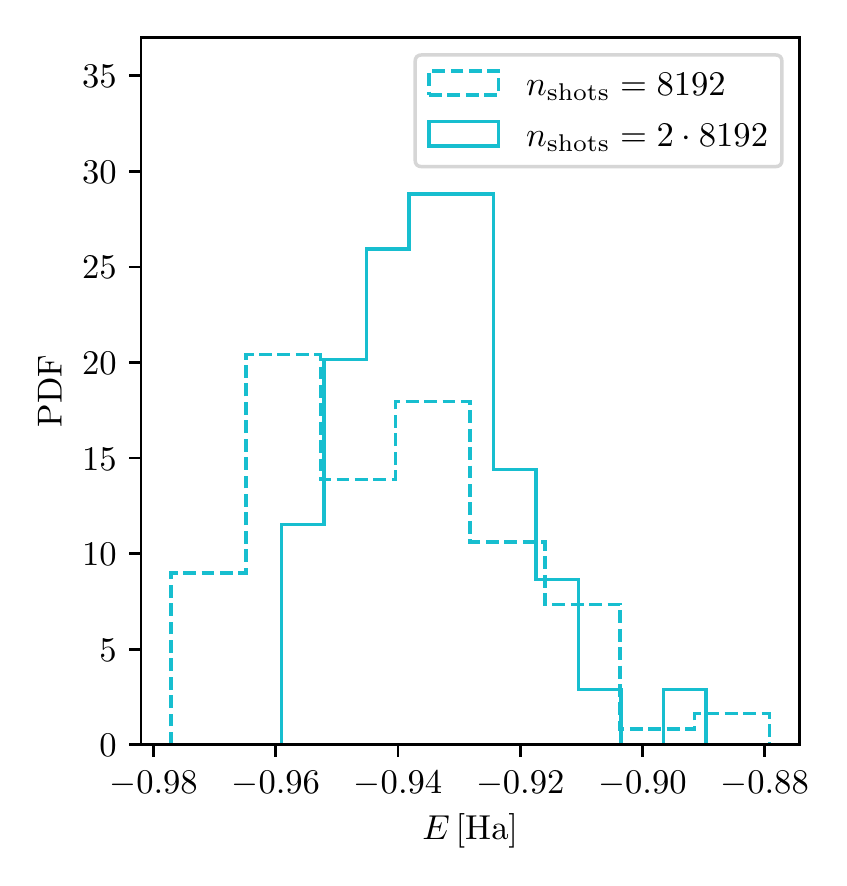}
    \caption{ Effect of increasing the number of measurements budget in the zero-noise-extrapolation error mitigation scheme. Shown is the distribution of the final ground state energy estimations obtained with 50 repetitions of the protocol (dashed line). Each value is obtained extrapolating from $4$ data points measured using $1,3,5$ and $7$ copies of each CZ gate. 
    The histogram acquired with double the number of shots ($2 \cdot 8192$) is reported with a solid line.}
    \label{fig:probing_different_zne_shemes}
\end{figure}

This section contains additional results for the experiments presented in the main text. In Fig.~\ref{fig:tet_h3_alma_lan} the result for the Lanczos method are shown for the parameters $a_0,a_1$ minimizing the energy expectation value for the dataset also presented in Fig.~3 in the main text. The results for the tetrahedron coincide with the ones given in the main text as the chosen parameters $a_0,a_1$ minimizing the energy yield an accuracy below $\sigma/J = 0.03$. In Fig.~\ref{fig:diss_h2_full} the results for the different Lanczos methods for the VQE results of $\mathrm{H}_2$ generated on four qubits are depicted. Also in this case the results for the parameters $a_0,a_1$ achieving the exact minimum of Eq.~\eqref{eq:lanczos} (blue) and the ones yielding a maximum standard error of $\SI{0.03}{Ha}$ (orange) almost coincide when the experimental estimates of $H,H^2,H^3$ are obtained with $n_\mathrm{shots}=8\cdot 10^5$. Finally, Fig.~\ref{fig:probing_different_zne_shemes} provides additional results for the comparison of the zero-noise extrapolation (ZNE) and the Lanczos error mitigation methods. In particular, we present the effect of increasing the total number of measurements in the ZNE approach.

\end{document}